\documentclass[aps,prb,twocolumn,amsmath,amssymb,superscriptaddress,floatfix,longbibliography]{revtex4-2}
\pdfoutput=1
\usepackage{mathrsfs}
\usepackage{amsmath}
\usepackage{mathtools}
\usepackage{braket}
\usepackage{amsfonts}
\usepackage{float}
\usepackage{graphicx} 
\usepackage[caption=false]{subfig}
\usepackage{bm} 
\usepackage{xcolor}
\usepackage{epstopdf}
\usepackage{comment}
\usepackage{soul}
\usepackage{multirow}
\usepackage{physics}
\usepackage[normalem]{ulem}
\usepackage{nicefrac, xfrac}
\usepackage{hhline}
\usepackage{simpler-wick}
\usepackage{titlesec}
\usepackage{cancel}
\usepackage{xr-hyper}
\usepackage{hyperref} 

\titleformat{\paragraph}
{\normalfont\itshape\normalsize\centering}{\theparagraph.}{1em}{}
\titlespacing*{\paragraph}
{0pt}{3.25ex plus 1ex minus .2ex}{1.5ex plus .2ex}

\newcommand\red[1]{\textcolor{red}{#1}}
\newcommand\blue[1]{\textcolor{blue}{#1}}

\newcommand{\Sum} [2] {\the\numexpr #1 + #2 \relax \\}

\begin{document}
\title{Order parameters and ground-state phase diagram of the interacting topological Su-Schrieffer-Heeger model with extended-range hoppings}

\author{ Tsz Hin Hui }
\affiliation{ \textit Department of Physics, City University of Hong Kong, Kowloon, Hong Kong, China}


\author{Pedro D. Sacramento}
\affiliation{ \textit CeFEMA, Instituto Superior T\'ecnico, Universidade de Lisboa, Av. Rovisco Pais, 1049-001 Lisboa, Portugal}

\author{ Wing Chi Yu }
\email{wingcyu@cityu.edu.hk}
\affiliation{ \textit Department of Physics, City University of Hong Kong, Kowloon, Hong Kong, China}

\date{\today}

\begin{abstract}
Topological insulators have attracted numerous attentions recent years, where the Su-Schrieffer-Heeger (SSH) model is one of the most studied models. While the interacting version of it has been explored recently, the interplay between interactions and long-range hoppings merit further investigations.
In this work, we uncover a rich phase diagram of the interacting SSH model with extended-range hoppings, in which it consists of several topological phases, two novel superconducting-like (SC-like) phases and five distinct charge-density-wave (CDW) phases. We substantiate that the SC-like and two CDW phases are direct consequences of imbalanced interactions and extended-range hoppings. We derive the order parameters (OPs) 
for each of the phases and verify them in large-system simulations, finding consistency with the entanglement entropy and the fidelity in capturing the phase transitions. In contrast to the non-interacting case where the favored hoppings are unidirectional in the topological phases,
the derived OPs suggest non-unidirectional hoppings are possible under the influence of interactions. 
\end{abstract}

\maketitle

\section{Introduction}

Symmetry plays a central role in modern physics. One of the most prominent example is Noether recognizing symmetries in the universe leads to conservation laws in nature. In many-body physics, Landau introduced one of the most elegant ideas that relates symmetries and phases of matter - phase transitions arise from symmetry breaking of the system, with emergence of an order parameter that captures the broken symmetry. However, since the seminal discovery of the integer quantum Hall effect \cite{wen1990,davidtongQHE}, researchers recognized the inadequacy in characterizing phases of matter entirely based on symmetries, leading to subsequent studies that developed the notion of topological phases of matter \cite{wen2017}. One example is the topological insulator \cite{Hasan2010,Qi2011}, which is an insulator in its bulk, but possesses conducting states on its surface.

One of the most extensively studied topological insulators is the Su-Schrieffer-Heeger (SSH) model \cite{ssh1979}, where the presence of edge states is guaranteed by the non-trivial topological structure in the wavefunction, quantified by a topological invariant $\mathbb{Z}$. The model has been realized in multiple experimental platforms \cite{Jalochowski2024,Cai2019,Mao2022}, and has been extensively studied in various settings, including non-equilibrium dynamics \cite{wong2024,vajna2015,yu2021,Sedlmayr2018,Zhang2022,Hou2022,li2022_pre,Rossi2022}, quantum batteries \cite{Zhou2026,Zhao2022}, disordered \cite{Hsu2020,song2014,lin2021,Perez-Gonzalez2019,Cinnirella2024}, non-Hermitian \cite{Medina-Guerra2025,Hajong2026,Youenn2023,Tzeng2021,Bacsi2021,Chang2020,Rafael2025,Chen2024,Nava2023,Arandes2025,Bissonnette2024}, and open dissipative quantum systems \cite{Nava2023,Dangel2018}. Several extensions of the model, such as multi-band versions \cite{maffei2018,Maslowski2020,Joshi2025}, higher spatial dimensions \cite{Sadrzadeh2021,Liu2017,Obana2019,liu2023_prb,xu2020,Zheng2019}, and longer range of hoppings \cite{maffei2018,wong2024,Rufo2019,Li2018,Li2014,Chang2025,Ghosh2023,Perez-Gonzalez2019,Cinnirella2024,Carmelo2025,Hajong2026,hui2026,Joshi2025}, have been considered in the literature. In particular, the inclusion of extended-range hoppings results in richer topological structure in the wavefunction characterized by higher winding numbers. This interest is driven, in part, by the increasing number of experimental platforms reporting rich topological phases induced by long-range hoppings \cite{Li2024,Liu2023_prapplied,Wang2023,Leefmans2022,he2026}, 
opening avenues to engineer novel artificial quantum phases of matter. While most of the above studies focus on the non-interacting SSH model, another line of research investigated the effect of uniform interactions \cite{melo2023,di2024,chen2018,sirker2014,marques2017,yahyavi2018,jin2023,matveeva2024,wagner2023,mondal2021,Barbiero2018,manmana2012,Lessnich2021}. However, only few studies have considered the case where the inter- and intra-cell interactions are imbalanced \cite{yu2016,yu2021,zhou2023,Marashli2025,Zhang2020,Nersesyan2020,Junemann2017}, and their interplay with the extended-range hoppings remains to be addressed.

One of the major challenges in characterizing the ground-state phase diagram of the interacting extended SSH (ESSH) model arises from the absence of well-established order parameters (OPs). In this work, we apply the scheme proposed in Ref. \cite{hui2026} to construct the OPs of various phases in the model. With the derived OPs, we identify a rich phase diagram, hosting ten phases. Remarkably, a superconducting-like (SC-like) phase is discovered in this model, which, to the best of our knowledge, has not been reported in previous literature except those studies that explicitly incorporate Peierls electron-phonon couplings \cite{Sous2018,Nocera2021}. The SC-like phase exhibits quasi-long-range order, as expected from the Mermin-Wagner theorem. Additionally, five types of charge density wave (CDW) are identified, in which two of them are a direct consequence of the interplay between extended-range hoppings and imbalanced interactions.

This paper is organized as follows. Section \ref{sec:review} reviews the pre-requisites for the paper, including an introduction of the SSH model and the general framework for constructing the OPs. Section \ref{sec:interactingSSH} addresses the interacting ESSH model, wherein Sec. \ref{sec:limiting_case} illustrates the possible phases under different limiting cases, while Sec. \ref{sec:OP_interacting} extends the scheme for OP constructions to various topological phases, as well as discusses the resulting phase diagrams in finite systems. The justification of the OPs in larger system is given in Sec. \ref{sec:DMRG_result}. Finally, a conclusion is given in Sec. \ref{sec:conclusion}.

\section{Review the pre-requisites}
\label{sec:review}

\subsection{The non-interacting SSH model}
\label{sec:review_ssh}

The Hamiltonian of the original SSH model reads:
\begin{align}
    H_\text{\cancel{o}SSH} &= \sum_{j = 1}^{N} ( t_ac_{j,A}^\dagger c_{j,B} + t_bc_{j,B}^\dagger c_{j + 1,A} + \text{h.c.} ),
    \label{eqn:H_oSSH}
\end{align}
where $N$ is the number of unit cells, \textit{A} and \textit{B} denote the two sublattice degrees of freedom. The model belongs to the BDI symmetry class \cite{ryu2010} associated with a topological invariant $\mathbb{Z}$. The topological invariant is identified to be the winding number $\mathcal{W}$ defined as
\begin{equation}
    \label{eqn:winding}
       \mathcal{W} = \frac{1}{2\pi}\int_{0}^{2\pi} {}_k\bra{-}\nabla_k\ket{-}_k dk  \ ,
\end{equation}
where $\ket{-}_k$ is the lower band eigenstate in the $k$-th mode. The integrand is $\frac{\partial\phi_k}{\partial k}$, where $\phi_k=\text{arg}(d^x_k+id^y_k)$ with $d^x_k = t_a + t_b\cos k $ and
$d^y_k = t_b \sin k$. When $\bigl|\frac{t_a}{t_b}\bigl|>1 $, $\phi_k$ is continuous, hence, $\mathcal{W}=0$ by fundamental theorem of calculus. The system is topologically trivial. However, $\phi_k$ is discontinuous (2 discontinuities in the $1^{\text{st}}$ Brillouin zone) for $\bigl|\frac{t_a}{t_b}\bigl|<1$. Using residue theorem, one finds $\mathcal{W}=1$ and the system is in the topologically non-trivial phase. 

The original SSH model only incorporates nearest-neighbor hoppings. To realize a richer topological structure in the ground state, it is necessary to include longer-range hoppings. For example, we may consider an ESSH model with the Hamiltonian \cite{wong2024,maffei2018}
\begin{align}
    H_\text{ESSH} &= \sum_{j = 1}^{N} ( t_ac_{j,A}^\dagger c_{j,B} + t_bc_{j,B}^\dagger c_{j + 1,A} \nonumber \\
    & \qquad + t_cc_{j,A}^\dagger c_{j + 1,B} + t_dc_{j,B}^\dagger c_{j + 2,A} + \text{h.c.} ).
    \label{eqn:H_ESSH}
\end{align}
The parameters $t_c$ and $t_d$ are the next-nearest neighbor hoppings which preserves the sublattice symmetry. Upon inclusion of the $t_c$ and $t_d$ terms, phases with $\mathcal{W}=-1,2$ are possible. In the following, we denote $\mathcal{W}=m$ as $\mathcal{W}_m$. From Ref. \cite{hui2026,wong2024,vajna2015}, we know that when $t_a/t_b/t_c/t_d$ dominates over the other hopping parameters, then the system is in the $\mathcal{W}_{0}/\mathcal{W}_{1}/\mathcal{W}_{2}/\mathcal{W}_{-1}$ phase.

One can go beyond next-nearest-neighbor hopping, and consider a general form of the non-interacting SSH model:
\begin{align}
    H = \sum_{j = 1}^{N} \sum_{i=-\gamma}^{\gamma}( t_i^{}c_{j,B}^\dagger c_{j+i,A}^{}  + \text{h.c.} ).
    \label{eqn:H_general}
\end{align}
 The notation $t_i$ in Eq. (\ref{eqn:H_general}) exposes the underlying topological phase naturally. In particular, when $t_m\gg t_l, \ \forall \ \ l\neq m$ and $l,m\in [-\gamma,\gamma]$, then the system is in the $\mathcal{W}_m$ phase. For the original SSH model, only $t_0\equiv t_a$ and $t_1\equiv t_b$ are finite, and all other hopping parameters are zero.

In the following, we will mainly focus on the Hamiltonian in Eq. (\ref{eqn:H_ESSH}). Periodic boundary condition (PBC) and the half-filling case are considered unless otherwise specified.

\subsection{The OP construction scheme}
\label{sec:review_op}
In this section, we review the scheme of constructing the OPs in Ref. \cite{hui2026}. Any many-body ground state can be written as:
\begin{align}
    \ket{\text{GS}} = \sum_{n_1{}_, n_2{}_,\cdots{}_, n_L}\xi_{n_1 n_2\cdots n_L}\ket{n_1 n_2\cdots n_L}.
    \label{eqn:fock_scheme}
\end{align}
The fundamental idea of the scheme is to extract the generic pattern in the dominant Fock states (i.e. Fock state $\ket{n_1 n_2\cdots n_L}$ with large $|\xi_{n_1 n_2\cdots n_L}|$) of the ground state in Eq. (\ref{eqn:fock_scheme}) to construct the OP. In particular, one has to analyze the Fock states that have dominant weights to infer the form of $\hat{O}$ that gives $\hat{O}\ket{n_1 n_2\cdots n_L}\neq 0$ on the dominant Fock states $\ket{n_1 n_2\cdots n_L}$, where $\hat{O}$ is the desired OP. As compared to existing approaches, our scheme is non-variational and not subsystem dependent. The construction process also allows researchers to gain immediate insight into the physical behavior of the system. 

From Ref. \cite{hui2026}, we found that, for a particular winding number phase $\mathcal{W}_m$ in Eq. (\ref{eqn:H_general}), the non-interacting ground state that lives deep inside the phase comprises all Fock states that have pairs of 1 and 0 across all sites connected by $t_m$. These Fock states have approximately equal weight in $|\xi_j|$ in Eq. (\ref{eqn:fock_scheme}), where $j$ is the integer corresponding to the binary representation of the Fock state. The number of these Fock states are $2^N$. For example, consider a system of $N=8$, the ground state deep inside the $\mathcal{W}_0$ phase has the following Fock states dominated:
\begin{equation}
 2^N\left\{ 
\begin{aligned}
&\quad |\widehat{1 0}\ \widehat{1 0}\ \widehat{1 0}\ \widehat{1 0}\ \widehat{1 0}\ \widehat{1 0}\ \widehat{1 0}\ \widehat{1 0} \rangle, \\
&\quad |\widehat{1 0}\ \widehat{1 0}\ \widehat{1 0}\ \widehat{1 0}\ \widehat{1 0}\ \widehat{1 0}\ \widehat{1 0}\ \widehat{0 1} \rangle, \\
&\quad |\widehat{1 0}\ \widehat{1 0}\ \widehat{1 0}\ \widehat{1 0}\ \widehat{1 0}\ \widehat{1 0}\ \widehat{0 1}\ \widehat{0 1} \rangle, \\
&\quad |\widehat{1 0}\ \widehat{1 0}\ \widehat{1 0}\ \widehat{1 0}\ \widehat{1 0}\ \widehat{0 1}\ \widehat{1 0}\ \widehat{0 1} \rangle, \\
&\quad |\widehat{1 0}\ \widehat{1 0}\ \widehat{1 0}\ \widehat{1 0}\ \widehat{1 0}\ \widehat{0 1}\ \widehat{0 1}\ \widehat{0 1} \rangle, \\
&\quad |\widehat{1 0}\ \widehat{1 0}\ \widehat{1 0}\ \widehat{1 0}\ \widehat{0 1}\ \widehat{1 0}\ \widehat{1 0}\ \widehat{0 1} \rangle, \\
&\quad |\widehat{1 0}\ \widehat{1 0}\ \widehat{1 0}\ \widehat{1 0}\ \widehat{0 1}\ \widehat{1 0}\ \widehat{0 1}\ \widehat{0 1} \rangle, \\
&\quad |\widehat{1 0}\ \widehat{1 0}\ \widehat{1 0}\ \widehat{0 1}\ \widehat{1 0}\ \widehat{0 1}\ \widehat{1 0}\ \widehat{0 1} \rangle, \\
&\quad |\widehat{1 0}\ \widehat{1 0}\ \widehat{0 1}\ \widehat{1 0}\ \widehat{1 0}\ \widehat{0 1}\ \widehat{1 0}\ \widehat{0 1} \rangle, \\
&\qquad \qquad \quad \vdots
\end{aligned}
\right.
\label{eqn:list_of_w0_fock}
\end{equation}
All the Fock states above are expressed in the form of $|\widehat{n_{1,A}^{} n_{1,B}^{}}\ \ \widehat{n_{2,A}^{} n_{2,B}^{}} \ \cdots \  \widehat{n_{8,A}^{} n_{8,B}^{}}\rangle$ and the hat above denotes a unit cell, 0 and 1 indicate the occupancy at each site. The key idea is that all the Fock states always have a pair of 0 and 1 at the sites $(j,A)$ and $(j,B)$, for all the cell $j$. Consequently, these sites favor the $t_a$ hopping. Therefore, the ground state of the $\mathcal{W}_0$ phase is approximately a superposition of any Fock state with the $t_a$ hopping present throughout the entire chain. 

The above observation plays a major role in the construction of the OP, in which we found that 
\begin{align}
O_{\mathcal{W}_0}^{3,\{/\}}&=c^\dag_{j,B}c_{j,A}^{}c^\dag_{j+1,B}c_{j+1,A}^{}c^\dag_{j+2,B}c_{j+2,A}^{}\ ,\\
&=c^\dag_{1B}c_{1A}^{}c^\dag_{2B}c_{2A}^{}c^\dag_{3B}c_{3A}^{} \ \ \ \ \ \ \text{if }\  j=1,
\label{eqn:w0_3}
\end{align}
where the first superscript denotes the presence of 3 hopping terms in this OP, while the second superscript $\{/\}$ will be explained later. The value of $j$ does not matter, since we impose PBC. Eventually, we found out that the OP of the non-interacting $\mathcal{W}_{0}$ phase is:
\begin{align}
\label{eqn:w0_non-interacting}
O_{\mathcal{W}_{0}} = 2^{n-1} \left(\prod_{j=1}^n c^\dag_{j,B}c_{j,A}^{} + \text{h.c.}\right).
\end{align}
More generally, i.e. for any phase in Eq. (\ref{eqn:H_general}),  we have:
\begin{align}
\label{eqn:wm_non-interacting}
O_{\mathcal{W}_m}&=  2^{n-1} \left( \prod_{j=1}^n c^\dag_{j+m,A}c_{j,B}^{} + \text{h.c.}\right)\  \text{for $m\in \mathbb{Z}$}.
\end{align}
The $n$ is interpreted as the number of hoppings in the OP. As noted in Ref. \cite{hui2026}, $n$ have to be chosen to grow linearly with $N$. Thus, when $N\rightarrow\infty$, this OP naturally involve multi-site hoppings and thus are non-local in real space, consistent with the momentum space defined winding numbers. However, as we will see later, the OPs for the non-interacting ESSH have to be refined when applying to the interacting model.

\section{The interacting ESSH model}
\label{sec:interactingSSH}
The Hamiltonian of the ESSH model with interactions is given by
\begin{align}
    H = H_\text{ESSH} + U\sum_j n_{j,A}n_{j,B} + V\sum_j n_{j,B}n_{j+1,A}\ ,
    \label{eqn:ISSH}
\end{align}
where $H_\text{ESSH}$ is defined in Eq. (\ref{eqn:H_ESSH}). To benchmark with the results in the interacting nearest-neighbor SSH model in Ref. \cite{yu2016}, we parametrize $t_a=-(1+\eta)$ and $t_b=-(1-\eta)$ with $\eta\in(-1,1)$ in this section unless otherwise specified. We first provide a qualitative discussion of the possible phases existing in the model by considering various limiting cases of ($U, V, \eta, t_c, t_d$) in Sec. \ref{sec:limiting_case}. Then, we demonstrate how to determine the OPs for certain phases in Sec. \ref{sec:OP_interacting}. Finally, we justify the results in large systems using density matrix renormalization group (DMRG) simulations in Sec. \ref{sec:DMRG_result}. 

\begin{figure}
    \centering
\begin{tikzpicture}
\centering
\draw[->] (1,0) -- (4,0) node[anchor=north west] {U};
\draw[-] (-3.3,0) -- (-3,0);
\draw[->] (0,1) -- (0,4) node[anchor=south east] {V};
\draw[-] (0,-3.3) -- (0,-3);
\draw (-3.3,3.3) -- (-3.3,-3.3) -- (3.3,-3.3) -- (3.3,3.3) -- (-3.3,3.3);
\draw (-1,1) -- (-1,-1) -- (1,-1) -- (1,1) -- (-1,1);
\draw[red] (-3,0) parabola (-1,1);
\draw[red] (-3,0) parabola (-1,-1);
\draw[blue] (1,-1) parabola (0,-3);
\draw[blue] (-1,-1) parabola (0,-3);
\node at (3.3,3.5) {$(+\infty,+\infty)$};
\node at (-3.3,-3.5) {$(-\infty,-\infty)$};
\node at (3.3,-3.5) {$(+\infty,-\infty)$};
\node at (-3.3,3.5) {$(-\infty,+\infty)$};
\filldraw[gray] (-3.3,3.3) circle (2pt);
\filldraw[gray] (-3.3,-3.3) circle (2pt);
\filldraw[gray] (3.3,-3.3) circle (2pt);
\filldraw[gray] (3.3,3.3) circle (2pt);
\node[align=left] at (0,0) {$\mathcal{W}_0/\mathcal{W}_1/$ \\ $\textcolor{red}{\mathcal{W}_2}/\textcolor{blue}{\mathcal{W}_{-1}}$ } ;
\node[align=left] at (-1.7,0) {\red{CDW-4A}};
\node[align=left] at (0,-1.2) {\blue{CDW-4B}};
\node[align=left] at (-2.5,2.5) {CDW-2A};
\node[align=left] at (-2.5,-2.5) {PS};
\node[align=left] at (2.5,2.5) {CDW-1A};
\node[align=left] at (2.5,-2.5) {CDW-2B};
\node[align=left] at (2,1.3) {$\mathcal{W}_0$};
\filldraw[gray] (3.3,0) circle (2pt);
\node at (3.85,0.25) {$(+\infty,0)$};
\draw[->] (2,1.2) -- (3.3,0); 
\draw[->] (2,1.2) -- (2.8,0); 
\draw[->] (2,1.2) -- (2.3,0); 
\draw[->] (2,1.2) -- (1.8,0); 
\draw[->] (2,1.2) -- (1.3,0); 
\node[align=left] at (1.5,1.9) {$\mathcal{W}_1$};
\node at (0.6,3.6) {$(0,+\infty)$};
\filldraw[gray] (0,3.3) circle (2pt);
\draw[->] (1.3,1.9) -- (0,3.3); 
\draw[->] (1.3,1.9) -- (0,2.8); 
\draw[->] (1.3,1.9) -- (0,2.3); 
\draw[->] (1.3,1.9) -- (0,1.8); 
\draw[->] (1.3,1.9) -- (0,1.3); 
\end{tikzpicture}
\caption{Schematic diagram illustrating possible phases in various limits of $(U,V)$. In the strong interacting regime, for any finite $\eta$, $t_c$ and $t_d$, the four phases reside at the four corners are phase separation (PS) for $(U,V)\rightarrow (-\infty,-\infty)$, and different types of charge density waves (CDWs), as detailed in the main text, for $(\pm\infty,\mp\infty)$ and $(+\infty,+\infty)$.  Along the line $(U,V)\approx (\lambda,0)$ and $(0,\lambda)$, for sufficiently large $\lambda>0$, there exists a $\mathcal{W}_0$ and a $\mathcal{W}_1$ phase, respectively, as indicated by the arrows. The central square represents the neighborhood of $(U,V)\approx(0,0)$, which admits four possibilities depending on the values of $\eta$, $t_c$ and $t_d$. For $|t_c|$ and $|t_d|$ sufficiently smaller than $|t_a|$ and $|t_b|$, as parametrized by $|\eta|$, the phase is $\mathcal{W}_0$ when $\eta>0$ and $\mathcal{W}_1$ when $\eta<0$. For $t_c=0$ and $|t_d|\gg |t_a|,|t_b| $, the central region is the $\mathcal{W}_2$ phase, along with the appearance of a CDW-4A phase as indicated in red. 
Similarly, for $t_d=0$ and $|t_c|\gg |t_a|,|t_b| $, the central region is the $\mathcal{W}_{-1}$ phase, along with the appearance of a CDW-4B phase as indicated in blue.
\label{fig:schematic_phasediagram}
}                                                                                 
\label{draw:schematic_UVlimit}
\end{figure}
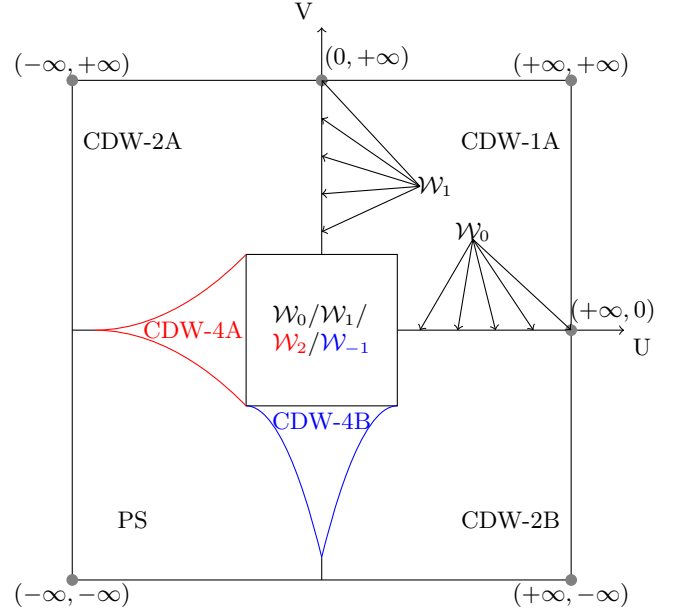

\subsection{Different limiting cases of  $(U,V,\eta,t_c,t_d)$}
\label{sec:limiting_case}
This subsection discuss certain possible phases that emerge under different limiting cases. The sub-sub-sections (\ref{subsubsec:CDW}-\ref{sec:interacting_limit_deep_w0_w1}) consider the case where the interaction strength ($|U|$ or $|V|$) is much larger than the hopping parameters ($|t_a|$, $|t_b|$, $|t_c|$, and $|t_d|$). In such a case, the values of the individual hopping parameters are unimportant in determining the phase. On the other hand, sub-sub-sections \ref{subsubsec:noninteracting_limit}-\ref{sec:limiting_case_CDW4A4B} consider scenarios where the relative values of $\eta, t_c, t_d$ becomes important in determining the phase.
A schematic diagram of the possible phases is summarized in Fig. \ref{draw:schematic_UVlimit}. 

\subsubsection{$(U,V)\rightarrow(-\infty,\infty)$ for any finite $(\eta,t_c,t_d)$}
\label{subsubsec:CDW}
The top-left phase in Fig. \ref{draw:schematic_UVlimit} is a CDW, which has periodicity of two unit cells. Taking an 8-unit cell system as an example, the dominate Fock state in the ground state is
\begin{eqnarray}
\label{Fock:cdw2a}
    |\widehat{11}\ \widehat{00}\ \widehat{11}\ \widehat{00}\ \widehat{11}\ \widehat{00}\ \widehat{11}\ \widehat{00}\rangle
\end{eqnarray}
and its global translation by one unit cell. Physically, for a strongly repulsive $V$, i.e. $V\rightarrow\infty$, the system avoids simultaneous occupancy across the $(j,B)$-$(j+1,A)$ sites. On the other hand, with $U\rightarrow-\infty$, the system still favors simultaneous occupancy across the $(j,A)$-$(j,B)$ sites. The only configurations that favors these two conditions are Eq. (\ref{Fock:cdw2a}) and its global translation.

From Ref. \cite{yu2016}, the OP takes the form
\begin{eqnarray}
    \label{op:cdw2a}
    O_{\text{CDW2A}}^{}(j)=\frac{1}{2}\bigl(n_{j,A}^{}+n_{j,B}^{}-n^{}_{j+1,A}-n^{}_{j+1,B}\bigl).\quad
\end{eqnarray}
As we will see, several types of CDW can emerge in this model. We denote them as ``CDW-mG'', where ``m'' is the period (in unit cells) of the wave, and G can be either A or B, indicating the starting site of the periodicity. For example, CDW-2A, as manifested by the Fock state in Eq. (\ref{Fock:cdw2a}), has a period of 2 unit cells, and corresponds to the ``1100'' wave pattern starting on the A site.

\subsubsection{$(U,V)\rightarrow(-\infty,-\infty)$ for any finite $(\eta,t_c,t_d)$}
\label{sec:limitingcase_ps}
There is a phase separation (PS) phase, as shown in the bottom-left corner in Fig. \ref{draw:schematic_UVlimit}, in this limit. In this phase, the spinless fermions cluster into a single high density domain, as evident by the dominant Fock states in the many-body ground state, namely,
\begin{align}
\label{Fock:ps}
    &|\widehat{11}\ \widehat{11}\ \widehat{11}\ \widehat{11}\ \widehat{00}\ \widehat{00}\ \widehat{00}\ \widehat{00}\rangle, \\
\label{Fock:ps2}
    &|\widehat{01}\ \widehat{11}\ \widehat{11}\ \widehat{11}\ \widehat{10}\ \widehat{00}\ \widehat{00}\ \widehat{00}\rangle. 
\end{align}
Each of the above represents a family of Fock states that are equivalent up to a global translation of one unit cell, and their particle-hole  counterparts obtained by mapping $1\rightarrow 0$ and $0\rightarrow 1$ on all sites. For instance, Eq. (\ref{Fock:ps}) represents the eight Fock states
\begin{gather*}
|\widehat{11}\ \widehat{11}\ \widehat{11}\ \widehat{11}\ \widehat{00}\ \widehat{00}\ \widehat{00}\ \widehat{00}\rangle, \ \ \ 
|\widehat{00}\ \widehat{00}\ \widehat{00}\ \widehat{00}\ \widehat{11}\ \widehat{11}\ \widehat{11}\ \widehat{11}\rangle, \\ 
|\widehat{00}\ \widehat{11}\ \widehat{11}\ \widehat{11}\ \widehat{11}\ \widehat{00}\ \widehat{00}\ \widehat{00}\rangle, \ \ \ 
|\widehat{11}\ \widehat{00}\ \widehat{00}\ \widehat{00}\ \widehat{00}\ \widehat{11}\ \widehat{11}\ \widehat{11}\rangle,\\ 
|\widehat{00}\ \widehat{00}\ \widehat{11}\ \widehat{11}\ \widehat{11}\ \widehat{11}\ \widehat{00}\ \widehat{00}\rangle, \ \ \ 
|\widehat{11}\ \widehat{11}\ \widehat{00}\ \widehat{00}\ \widehat{00}\ \widehat{00}\ \widehat{11}\ \widehat{11}\rangle,\\ 
|\widehat{00}\ \widehat{00}\ \widehat{00}\ \widehat{11}\ \widehat{11}\ \widehat{11}\ \widehat{11} \ \widehat{00}\rangle, \ \ \ 
|\widehat{11}\ \widehat{11}\ \widehat{11}\ \widehat{00}\ \widehat{00}\ \widehat{00}\ \widehat{00} \ \widehat{11}\rangle, 
\end{gather*}
where the right column is the particle-hole counterpart of the left. All of the eight Fock states above share the same $|\xi_i|^2$ in Eq. (\ref{eqn:fock_scheme}) due to translational symmetry. Similar applies to Eq. (\ref{Fock:ps2}) and the subsequent discussions unless otherwise specified.

Physically, in the strongly attractive regime where both $U$ and $V$ tend to $-\infty$, the system favors simultaneous occupancy across the $(j,A)$-$(j,B)$ sites, and the $(j,B)$-$(j+1,A)$ sites. The only way to satisfy this throughout the lattice in half-filling is for the spinless fermions cluster into contiguous block, which is precisely the PS phase configuration.

One may view the PS phase as a CDW whose wavelength spans the whole system. Referencing to the OP in Eq. (\ref{op:cdw2a}), we
propose the OP for the PS phase as
\begin{eqnarray}
\label{op:ps}
O_{\text{PS}}^{}(j)=\frac{1}{2}\bigl(n_{j,A}^{}+n^{}_{j,B}-n^{}_{j+\frac{N}{2},A}-n^{}_{j+\frac{N}{2},B}\bigl).
\end{eqnarray}
Note that the OPs of the CDW family have four density operators, two of which have positive signs, and the remaining have negative signs (see Eq. (\ref{op:cdw2a}) and (\ref{op:ps})). The rule of thumb to construct the OPs of this family is that the two density operators with positive (negative) signs have to be acted on the first two sites of ``1'' (``0'') in the dominant Fock states. Therefore, $O_{\rm{PS}}$ spans across half the wavelength of the PS phase. This rule of thumb also applies to all other CDW phases in this paper, see Eqs. (\ref{op:ps}),  (\ref{op:cdw4}), and (\ref{op:cdw4B}) for more evident example. We will verify the capability of these OPs to capture their respective phases in the later subsections.

\subsubsection{$(U,V)\rightarrow (\infty,-\infty)$ for any finite $(\eta, t_c,t_d)$}
\label{sec:limitingcase_CDW2B}
This phase locates at the bottom-right corner in Fig. \ref{draw:schematic_UVlimit}. The ground state of the system in this limit is mainly contributed by the Fock state
\begin{eqnarray}
\label{fock:cdw2b}
    |\widehat{1 \ 0}\widehat{0\ 1}\widehat{1\ 0}\widehat{0\ 1}\widehat{1\ 0}\widehat{0\ 1}\widehat{1\ 0}\widehat{0\ 1}\rangle
\end{eqnarray}
and its global translation by one unit cell. Similar to the configuration in Eq. (\ref{Fock:cdw2a}), these Fock states has a period of two unit cells, except that the wave ``1100'' now starts at site B while the one in Eq. (\ref{Fock:cdw2a}) starts at site A. 
Therefore, we label this phase as the CDW-2B phase. 
The OP of this phase has been obtained in Ref. \cite{yu2016} and it takes the form
\begin{equation}
    O_{\text{CDW2B}}^{}(j)=\frac{1}{2}\bigl(n_{j,B}^{}+n^{}_{j+1,A}-n^{}_{j+1,B}-n^{}_{j+2,A}\bigl).
    \label{op:cdw2b}
\end{equation}

\subsubsection{$(U,V)\rightarrow(\infty,\infty)$ for any finite $(\eta,t_c,t_d)$ }
This phase reside at the top-right corner in Fig. \ref{draw:schematic_UVlimit}, labeled as the CDW-1A phase (or CDW-1B, which are equivalent under our notation). The dominant Fock state in the ground state of this phase is 
\begin{eqnarray}
\label{Fock:cdw1a}
    |\widehat{10}\ \widehat{10}\ \widehat{10}\ \widehat{10}\ \widehat{10}\ \widehat{10}\ \widehat{10}\ \widehat{10}\rangle
\end{eqnarray}
and its particle-hole counterpart obtained by mapping $1\rightarrow0$ and $0\rightarrow1$ for each site. This phase has a period of 1 unit cell, and hence, is termed as CDW-1A. Literatures refers this CDW-1A phase as the topologically trivial Mott insulting phase \cite{melo2023}.

Referencing to Ref. \cite{yu2016}, and the OPs in Eq. (\ref{op:cdw2a}) and Eq. (\ref{op:cdw2b}), it is natural to write down the OP of the CDW-1A phase as
\begin{equation}
\label{op:cdw1a}
    O_{\text{CDW1A}}^{}(j)=\frac{1}{2}\bigl(n_{j,A}^{}+n^{}_{j+1,A}-n^{}_{j,B}-n^{}_{j+1,B}\bigl).
\end{equation}



\subsubsection{$(U,V)\rightarrow (\infty,0)$ or $(U,V)\rightarrow (0,\infty)$ for any \\finite $(\eta, t_c,t_d)$}
\label{sec:interacting_limit_deep_w0_w1}
These two cases correspond to the positive $x$- and $y$-axis in Fig. \ref{draw:schematic_UVlimit}, respectively, as indicated by the arrows in the plot. Typically, interactions destroy topology \cite{Shao2021}. However, in the interacting ESSH model, we can have topology being enhanced as certain interactions is turned on. Along the line $U=0$ with $V\rightarrow\infty$, the feature of the topological $\mathcal{W}_{1}$ phase gets stronger. On the other hand, if $U\rightarrow\infty$ while keeping $V=0$, the system is always in a topologically trivial $\mathcal{W}_{0}$ phase. 

To provide insights to the above claim, we consider $(U,V)\rightarrow (\infty,0)$, which implies strong repulsive interactions between the two sites within a unit cell. The spinless fermions will avoid simultaneously occupying the $(j,A)$ and $(j,B)$ sites for all $j$. In turn, Fock states that have a pair of 0 and 1 at the $(j,A)$ and $(j,B)$ sites, for all $j$, should have dominant contribution to the ground state, which is a characteristic of a strong $\mathcal{W}_0$ phase. 
If we introduce a small but finite $V$, i.e. departing from the $U$-axis, the $\mathcal W_0$ phase feature will be weakened. More concrete example will be presented and discussed in the next section.
Following similar logic, one can argue that $(U,V)\rightarrow (0,\infty)$ corresponds to deep inside the $\mathcal{W}_1$ phase and we shall not repeat the argument here.

\subsubsection{$(U,V)\approx(0,0)$}
\label{subsubsec:noninteracting_limit}
This case corresponds to the cental square box in Fig. \ref{draw:schematic_UVlimit}. In this non-interacting limit, the phase that would exist depends entirely on the values of $\eta$, $t_c$, and $t_d$. For example, if $(t_c,t_d)=(0,0)$, the model reduces to the nearest-neighbor hopping interacting SSH model, the phase will be $\mathcal{W}_0$ and $\mathcal{W}_1$ for $\eta>0$ and $\eta<0$, respectively. An example of the phase diagram for $\eta=0.6$ and $\eta=-0.6$ can be found in Ref. \cite{yu2016}. However, as we increase the magnitude of $t_c$ or $t_d$, this region might become a $\mathcal{W}_{-1}$ or a $\mathcal{W}_2$ phase, respectively. We will show this is in fact the case in Sec. \ref{sec:OP_interacting}.

\subsubsection{$(U\ll 0,V\approx 0, t_c = 0, t_d < 0)$ or $(U\approx 0,V\ll 0, t_c<0, t_d = 0)$ for any $\eta$ }
\label{sec:limiting_case_CDW4A4B}

These two cases correspond to the phase termed ``CDW-4A'' and `CDW-4B'' in Fig. \ref{draw:schematic_UVlimit}, respectively. As the name suggests, the dominant Fock state in the ground state of the ``CDW-4A'' phase is
\begin{eqnarray}
    \label{Fock:cdw4A}
    |\widehat{11}\ \widehat{11}\ \widehat{00}\ \widehat{00}\ \widehat{11}\ \widehat{11}\ \widehat{00}\ \widehat{00}\rangle, 
\end{eqnarray}
along with another three counterpart Fock states. On the other hand, the dominant Fock state in the ground state of the ``CDW-4B'' phase is
\begin{eqnarray}
    \label{Fock:cdw4B}
    |\widehat{11}\ \widehat{10}\ \widehat{00}\ \widehat{01}\ \widehat{11}\ \widehat{10}\ \widehat{00}\ \widehat{01}\rangle, 
\end{eqnarray}
and its three other counterparts. Only one of the two phases will appear on the U-V phase diagram in Fig. \ref{draw:schematic_UVlimit}, depending on the choice of $t_c$ and $t_d$.
If $\mathcal{W}_2$($\mathcal{W}_{-1})$ is the phase residing at the central region, which is the central square box in the figure, then ``CDW-4A'' (``CDW-4B'') will exist. 
These two phases are absent in the interacting SSH model with only nearest-neighbor hoppings \cite{yu2016}.

To illustrate the physical intuition behind the emergence of the CDW-4A or CDW-4B phases, let us consider the case of $t_d<0$. The same reasoning applies to the case of $t_c<0$. First, notice that when $U\ll0$ and $V=0$, the system favors to have the $(j,A)$ and the $(j,B)$ sites either being both occupied or both empty in the half-filling case. Two possible example configurations are those in Eq. (\ref{Fock:ps}) and in Eq. (\ref{Fock:cdw2a}). Under strongly attractive $U$, the effects of $t_a$ and $t_b$ are negligible. When $t_d<0$, the system further favors the $(j,B)$ and the $(j+2,A)$ sites having a 0 and 1 pair, yet neither Eq. (\ref{Fock:ps}) nor Eq. (\ref{Fock:cdw2a}) satisfy this condition. In order to satisfy this condition while keeping the system also favor the configurations when $U\ll0$ and $V=0$, the family of Fock states in Eq. (\ref{Fock:cdw4A}) is the only candidate. 

We make reference to the OPs of the CDW phases in Eqs. (\ref{op:cdw2a}), (\ref{op:cdw2b}), and (\ref{op:cdw1a}), and write down the OP for the CDW-4A phase as
\begin{eqnarray}
    \label{op:cdw4}
    O_{\text{CDW4A}}^{}(j)=\frac{1}{2}\bigl(n_{j,A}^{}+n_{j,B}^{}-n^{}_{j+2,A}-n^{}_{j+2,B}\bigl).\ \ \ \ \
\end{eqnarray}
Similarly, we can also write down the OP of the CDW-4B phase
\begin{eqnarray}
    \label{op:cdw4B}
    O_{\text{CDW4B}}^{}(j)=\frac{1}{2}\bigl(n_{j,B}^{}+n_{j+1,A}^{}-n^{}_{j+2,B}-n^{}_{j+3,A}\bigl). \quad \  
\end{eqnarray}

\begin{figure}
    \centering
    \includegraphics[width=8cm]{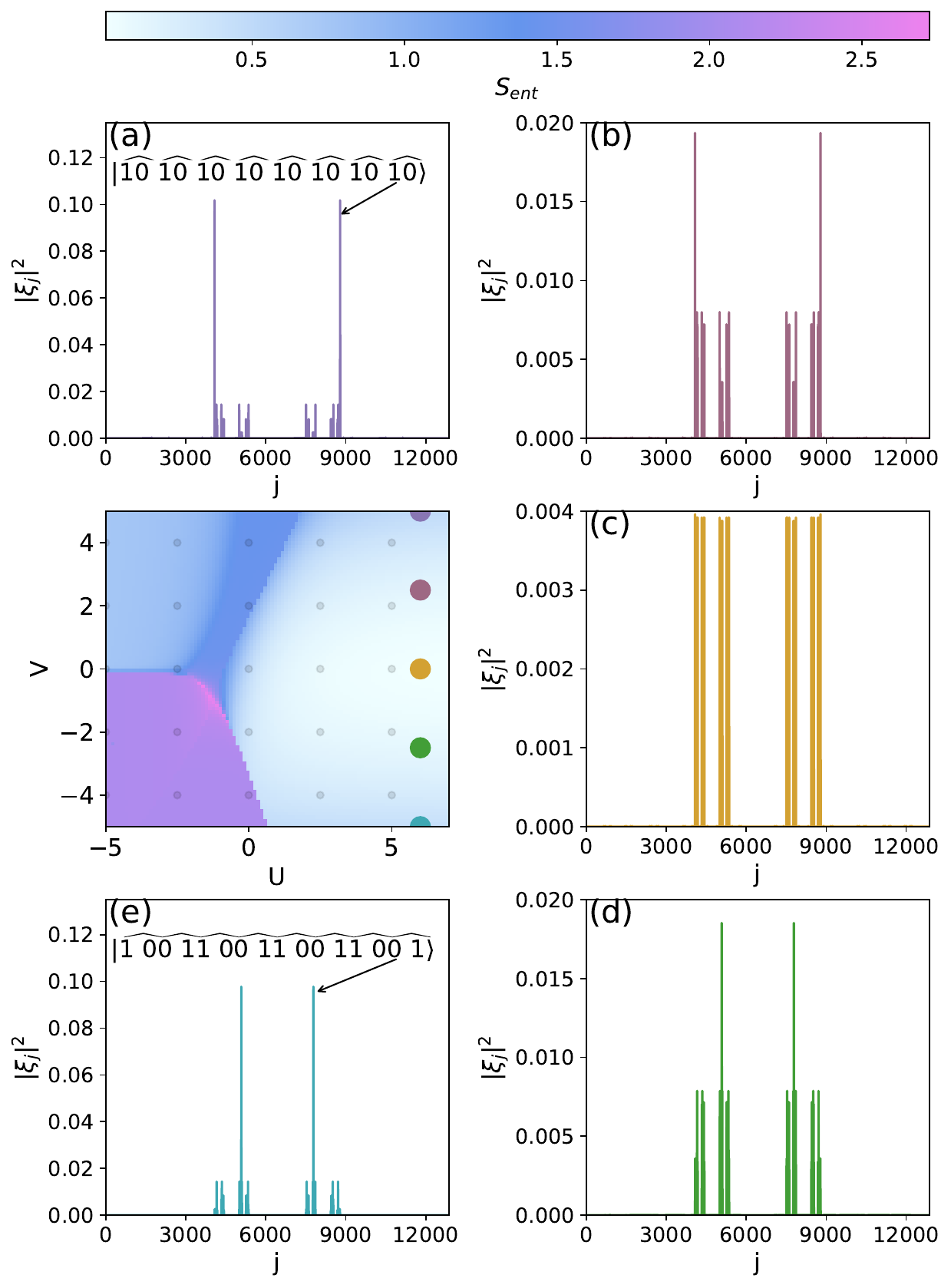}
    \caption{The squared modulus of the weight  $|\xi_j|^2$ (see Eq. (\ref{eqn:fock_scheme})) of the Fock states for ground states along $U=6$ with (a) $V=5$; (b) $V=2.5$; (c) $V=0$; (d) $V=-2.5$; (e) $V=-5$. The colormap shows the half-block entanglement entropy, with the circles marking the points at which the ground state is considered in panels (a-e) using the same color scheme. Here, $N=8$ and $(\eta,t_c,t_d)=(0.6,0,0)$.}
    \label{fig:pathFock}
\end{figure}

\subsection{OPs and estimated phase diagrams in small systems}
\label{sec:OP_interacting}

In this subsection, we apply the OPs above with some refinements to estimate the ground state phase diagram as a function of $U$ and $V$ for various $t_d\leqslant 0$ while fixing $(t_c,\eta)=(0,0.6)$. We first consider the case of $t_d=0$ where the results can be bench-marked with Ref. \cite{yu2016}. We mainly focus on finite-size systems, i.e. $N=8$, in this subsection, and justify the results in larger systems in Sec. \ref{sec:DMRG_result}.  

According to Ref. \cite{yu2016}, the interacting system exhibits four phases at $(\eta, t_c,t_d)=(0.6,0,0)$, namely, the $\mathcal{W}_0$, $\mathcal{W}_1$, PS, and CDW-2A phases (the last of which was referred to as the CDW phase in Ref. \cite{yu2016}). The OPs for the PS and CDW-2A phases are defined in Eq. (\ref{op:ps}) and Eq. (\ref{op:cdw2a}), respectively. For the non-interacting case, the OP for the $\mathcal{W}_1$ phase is:
\begin{align}
\label{eqn:w1_non-interacting}
O_{\mathcal{W}_1} &= 2^{n-1} \left(\prod_{j=1}^n c^\dag_{j+1,A}c_{j,B}^{} + \text{h.c.}\right) .
\end{align}
The OP for the $\mathcal{W}_0$ phase is in Eq. (\ref{eqn:w0_non-interacting}). These OPs [and also Eq. (\ref{eqn:wm_non-interacting})] are constructed from the product of unidirectional hopping operators. With interactions, however, the set of dominant Fock states no longer generally favors purely unidirectional hoppings. Consequently, the OPs must be refined to accurately capture the corresponding phases, as detailed below.

\subsubsection{Refinement of the $\mathcal{W}_m$ phases' OPs \\ in the interacting regime}

To illustrate the main idea, we here focus on the OPs in  Eqs. (\ref{eqn:w0_non-interacting}) and  (\ref{eqn:w1_non-interacting}) when $N=8$.
Figure \ref{fig:pathFock} displays the Fock states' probability distribution $\{|\xi_i|^2\}$ of the ground states with different $V$ along the $U=6$ path, where the $x$-axis is the integer corresponding to the binary representation of the Fock state, and the $y$-axis is the squared modulus of the corresponding weight ($|\xi_j|^2$). 
Note that Fig. \ref{fig:pathFock}(c) is qualitatively similar to the distribution of non-interacting ground state from deep inside the $\mathcal{W}_0$ phase \cite{hui2026}, and it has $2^N$ dominant Fock states. This supports the claim that the system lies deeper in the $\mathcal{W}_0$ phase as $(U,V)\rightarrow(\infty,0)$, as argued in Sec. \ref{sec:interacting_limit_deep_w0_w1}. As we introduce $V\neq 0$, the $\mathcal{W}_0$ characteristic is weakened, and the corresponding dominant Fock states no longer have approximately equal weights unlike in Fig. \ref{fig:pathFock}(c). The number of dominant Fock states become less than $2^N$.

Consider the case of $(\eta,t_c,t_d,U,V)=(0.6,0,0,6,5)$ in Fig. \ref{fig:pathFock}(a), there are 58 Fock states satisfying $|\xi_i|^2>0.003$ (the threshold is chosen on a case-by-case basis), and they are: 
\begin{eqnarray}
\label{idkfock1}
&|\widehat{1 0}\ \widehat{1 0}\ \widehat{1 0}\ \widehat{1 0}\ \widehat{1 0}\ \widehat{1 0}\ \widehat{1 0}\ \widehat{1 0}\rangle \ \  \ |\xi_i|=0.31905 , \\
\label{idkfock2}
&|\widehat{\underline{0 1}} \ \widehat{1 0} \ \widehat{1 0}\ \widehat{1 0}\ \widehat{1 0}\ \widehat{1 0}\ \widehat{1 0}\ \widehat{1 0}\rangle \ \ \ |\xi_i|=0.12012 , \\
\label{idkfock3}
&|\widehat{\underline{0 1}}\ \widehat{\underline{0 1}} \ \widehat{1 0}\ \widehat{1 0}\ \widehat{1 0}\ \widehat{1 0}\ \widehat{1 0}\ \widehat{1 0}\rangle \ \ \ |\xi_i|=0.09054 , \\
\label{idkfock4}
&|\widehat{\underline{0 1}}\ \widehat{\underline{0 1}} \ \widehat{\underline{0 1}}\ \widehat{1 0}\ \widehat{1 0}\ \widehat{1 0}\ \widehat{1 0}\ \widehat{1 0}\rangle \ \ \ |\xi_i|=0.08151 , \\
\label{idkfock5}
&|\widehat{\underline{0 1}}\ \widehat{\underline{0 1}} \ \widehat{\underline{0 1}}\ \widehat{\underline{0 1}}\ \widehat{1 0}\ \widehat{1 0}\ \widehat{1 0}\ \widehat{1 0}\rangle \ \ \ |\xi_i|=0.07924.
\end{eqnarray} 
Each of the above represents a set of Fock states that are equivalent up to a global translation by a unit cell, and also their Hermitian conjugate counterparts. The underline on certain cells denote a local discrepancy relative to
the remaining cells. For example, in the Fock state in Eq. (\ref{idkfock3}), there are two adjacent pairs of “01”, whereas the
other sites are “10”. We find that the above Fock states have configurations that favor $t_a$ hopping across all sites.

\begin{figure}
    \centering
    \includegraphics[width=8cm]{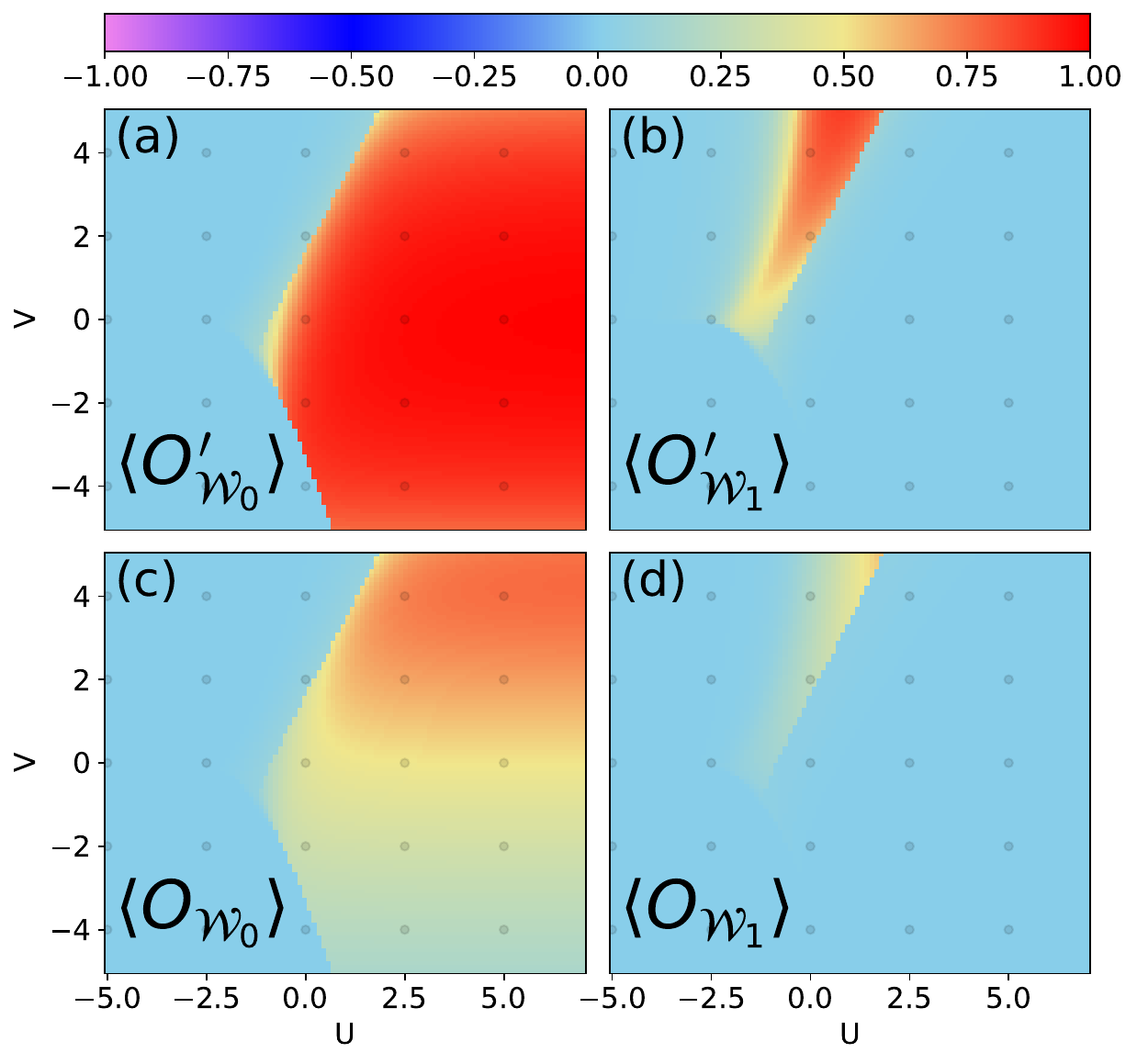}
    \caption{Ground-state expectation values of $O_{\mathcal{W}_0}'$, $O_{\mathcal{W}_1}'$,
    $O_{\mathcal{W}_0}$, and $O_{\mathcal{W}_1}$, defined in Eqs. (\ref{eqn:O^3_w0}), (\ref{eqn:O^3_w1}), (\ref{eqn:w0_non-interacting}) and (\ref{eqn:w1_non-interacting}), respectively. The number of hopping terms is set to $n=3$ for all the OPs. Here, $N=8$ and $(\eta,t_c,t_d)=(0.6,0,0)$. For comparison, see the phase diagram in Ref. \cite{yu2016}. }
    \label{fig:builidingInteractingOp}
\end{figure}

On the other hand, for the case of $(\eta,t_c,t_d,U,V)=(0.6,0,0,6,-5)$ in Fig. \ref{fig:pathFock}(e), the 58 Fock states with $|\xi_i|^2>0.003$ are: 
\begin{eqnarray}
\label{cdw2fock1}
&|\widehat{1 \ 0}\widehat{0\ 1}\widehat{1\ 0}\widehat{0\ 1}\widehat{1\ 0}\widehat{0\ 1}\widehat{1\ 0}\widehat{0\ 1}\rangle \ \ \ |\xi_i|=0.31265, \\
\label{cdw2fock2}
&|\widehat{1\ \underline{0}}\widehat{\underline{1}\ \underline{0}}\widehat{\underline{1}\ 0}\widehat{0\ 1}\widehat{1\ 0}\widehat{0\ 1}\widehat{1\ 0}\widehat{0\ 1}\rangle \ \ \ |\xi_i|=0.11986, \\
\label{cdw2fock3}
&|\widehat{1\ \underline{0}}\widehat{\underline{1}\ 0}\widehat{0\ \underline{1}}\widehat{\underline{0}\ 1}\widehat{1\ 0}\widehat{0\ 1}\widehat{1\ 0}\widehat{0\ 1}\rangle \ \ \ |\xi_i|=0.09161, \\
\label{cdw2fock4}
&|\widehat{1\ \underline{0}}\widehat{\underline{1}\ 0}\widehat{0\ 1}\widehat{1\ \underline{0}}\widehat{\underline{1}\ 0}\widehat{0\ 1}\widehat{1\ 0}\widehat{0\ 1}\rangle \ \ \ |\xi_i|=0.08308, \\
\label{cdw2fock5}
&|\widehat{1\ \underline{0}}\widehat{\underline{1}\ 0}\widehat{0\ 1}\widehat{1\ 0}\widehat{0\ \underline{1}}\widehat{\underline{0}\ 1}\widehat{1\ 0}\widehat{0\ 1}\rangle \ \ \ |\xi_i|=0.08095.
\end{eqnarray} 
Similar to the case of Eqs. (\ref{idkfock1}-\ref{idkfock5}), all the above Fock states favor $t_a$ hoppings. However, the dominant Fock state in Eq. (\ref{cdw2fock1}) favors $t_a$ hoppings in different directions, unlike that in Eq. (\ref{idkfock1}) where the $t_a$ hoppings are all in the same direction. When applying the OP [i.e. Eq. (\ref{eqn:w0_non-interacting})] with $n=3$ to the Fock state in Eq. (\ref{cdw2fock1}), it vanishes, unlike Eq. (\ref{idkfock1}), where it remains finite. Therefore, the sign of the interaction $V$ restricts the preferred Fock states, as clearly indicated in Fig. \ref{fig:pathFock}.

To find the OP, we have to revisit the non-interacting case. In fact, apart from Eq. (\ref{eqn:w0_3}), we can also construct seven other possibilities for the order parameter of the $\mathcal{W}_0$ phase, namely:
\begin{align}
\label{eqn:w0_3_1}
O_{\mathcal{W}_0}^{3,\{1\}}&=c^\dag_{1A}c_{1B}^{} c^\dag_{2B}c_{2A}^{}c^\dag_{3B}c_{3A}^{} \ ,\\
\label{eqn:w0_3_2}
O_{\mathcal{W}_0}^{3,\{2\}}&=c^\dag_{1B}c_{1A}^{}  c^\dag_{2A}c_{2B}^{} c^\dag_{3B}c_{3A}^{} \ ,\\
\label{eqn:w0_3_3}
O_{\mathcal{W}_0}^{3,\{3\}}&=c^\dag_{1B}c_{1A}^{}  c^\dag_{2B}c_{2A}^{} c^\dag_{3A}c_{3B}^{} \ ,\\
\label{eqn:w0_3_12}
O_{\mathcal{W}_0}^{3,\{1,2\}}&=c^\dag_{1A}c_{1B}^{}  c^\dag_{2A}c_{2B}^{}c^\dag_{3B}c_{3A}^{} \ ,\\
\label{eqn:w0_3_23}
O_{\mathcal{W}_0}^{3,\{2,3\}}&=c^\dag_{1B}c_{1A}^{}  c^\dag_{2A}c_{2B}^{} c^\dag_{3A}c_{3B}^{} \ ,\\
\label{eqn:w0_3_13}
O_{\mathcal{W}_0}^{3,\{1,3\}}&=c^\dag_{1A}c_{1B}^{}  c^\dag_{2B}c_{2A}^{} c^\dag_{3A}c_{3B}^{} \ ,\\
\label{eqn:w0_3_123}
O_{\mathcal{W}_0}^{3,\{1,2,3\}}&=c^\dag_{1A}c_{1B}^{}  c^\dag_{2A}c_{2B}^{} c^\dag_{3A}c_{3B}^{}\ ,
\end{align}
which all satisfy the requirement of having a 0 and 1 pair at the sites $(j,A)$ and $(j,B)$. Equations (\ref{eqn:w0_3_1}-\ref{eqn:w0_3_123}) follow from Eq. (\ref{eqn:w0_3}) by taking Hermitian conjugate locally for each hopping. The second superscript $\{...\}$ is a set denoting the location at which the local Hermitian conjugate is taken. Therefore, the superscript $\{/\}$ in Eq. (\ref{eqn:w0_3}) means that no local Hermitian conjugation is applied. All these seven possibilities should give larger expectation values for the ground state in the non-interacting $\mathcal{W}_0$ phase than in other phases. 

Returning to the problem we faced, one finds that $O_{\mathcal{W}_0}^{3,\{2\}}$ in Eq. (\ref{eqn:w0_3_2}) gives non-vanishing results when acting on the Fock state in Eq. (\ref{cdw2fock1}). Therefore, the OP of the interacting $\mathcal{W}_0$ phase must contain terms like Eq. (\ref{eqn:w0_3_2}). It turns out the OP for the interacting $\mathcal{W}_0$ phase is 
\begin{align}
    O_{\mathcal{W}_0}^{'3} = \sum_{\kappa\in\mathcal{P}({\{1,2,3\})}} O_{\mathcal{W}_0}^{3,\kappa}\ ,
\label{eqn:O^3_w0}
\end{align}
where $\mathcal{P}(\{1,2,3\})$ denotes the power set of $\{1,2,3\}$. This OP is a superposition of Eq. (\ref{eqn:w0_3}) and Eqs.(\ref{eqn:w0_3_1}-\ref{eqn:w0_3_123}). Since $O_{\mathcal{W}_0}^{'3}$ includes all possible combinations of hopping directions associated with $t_a$, it can automatically compensate for the tendency to favor certain Fock states induced by interactions. Figure \ref{fig:builidingInteractingOp}(a) shows the ground state expectation value of $O_{\mathcal{W}_0}^{'3}$. This OP not only has finite values in its own phase, but also attains similar values throughout the entire $\mathcal{W}_0$ phase, except near the phase boundary. For comparison,  the non-interacting case's OP in Eq. (\ref{eqn:w0_non-interacting}) is shown in Fig. \ref{fig:builidingInteractingOp}(c). Although it locates the phase boundary qualitatively well, its magnitude does not correctly reflect how deep the system is in the phase. In particular, as argued in  Sec. \ref{sec:interacting_limit_deep_w0_w1}, in the limit of $(U,V)\rightarrow(\infty,0)$ with all the hopping parameters being finite, the $\mathcal{W}_0$ phase character is the strongest and so the OP should take larger values than in the $V\ne0$ regime.

Similarly, for the $\mathcal{W}_1$ phase, one may consider the phase diagram at $\eta=-0.6$ and follow arguments analogous to those above to conclude that the appropriate OP is 
\begin{align}
\label{eqn:O^3_w1}
    O_{\mathcal{W}_1}^{'3} = \sum_{\kappa\in\mathcal{P}({\{1,2,3\})}} O_{\mathcal{W}_1}^{3,\kappa}\ ,
\end{align}
with $O_{\mathcal{W}_1}^{3,\{/\}}=c^\dag_{1,B}c_{2,A}^{}c^\dag_{2,B}c_{3,A}^{}c^\dag_{3,B}c_{4,A}^{}$. The corresponding result is shown in Fig. \ref{fig:builidingInteractingOp}(b), which has the desired properties as expected. The result of $O_{\mathcal{W}_1}$ is also shown in Fig. \ref{fig:builidingInteractingOp}(d) for comparison. When $t_d\ne 0$ ($t_c\ne 0$), the $\mathcal{W}_2$ ($\mathcal{W}_{-1}$) phase may emerge as discussed in previous sections. By the same reasoning, the OP for the $\mathcal{W}_{2}$ ($\mathcal{W}_{-1}$) phase should be $O_{\mathcal{W}_2}^{'3}$ ($O_{\mathcal{W}_{-1}}^{'3}$) in analogous to Eqs. (\ref{eqn:O^3_w0}) and (\ref{eqn:O^3_w1}). 

For completeness, we write down the OP for the interacting ground state in any winding number phase with $n$ hoppings:
\begin{align}
    \label{eqn:wm_combination}
    O_{\mathcal{W}_m}'&= \sum_{\kappa\in\mathcal{P}(\{ 1,2,\cdots,n\})} 
    O^{n,\kappa}_{\mathcal{W}_m} \ ,
\end{align}
where $\mathcal{P}(\{ 1,2,\cdots,n\})$ is the power set of $\{ 1,2,\cdots,n\}$, meaning we have $2^n$ terms in the above summation, and $O^{n,\kappa}_{\mathcal{W}_m}$ is defined as:
\begin{align}
O^{n,\kappa}_{\mathcal{W}_m}=    \prod_{j=1}^n c^\dag_{j+m,B}c_{j,A}^{}.
\end{align}
Equations (\ref{eqn:O^3_w0}) and (\ref{eqn:O^3_w1}) are recovered by setting $m=0$ and $m=1$, respectively, and $n=3$. A remark on the number of hoppings $n$ in the OPs in Eq. (\ref{eqn:wm_combination}). In the non-interacting case, we increase $n$ as the system size increases because the spinless fermions tend to delocalize in real space when only hopping terms are present in the Hamiltonian. In contrast,  
when interactions are present, they appear as diagonal terms in the real space Hamiltonian which promote localization of the particles. 
Therefore, in the interacting regime, it is necessary to truncate the number of hopping terms in the OPs. In the remaining discussion of the interacting ESSH model, we set $n=3$ in Eq. (\ref{eqn:wm_combination}) unless otherwise specified.

\begin{figure*}
\includegraphics[width=16cm]{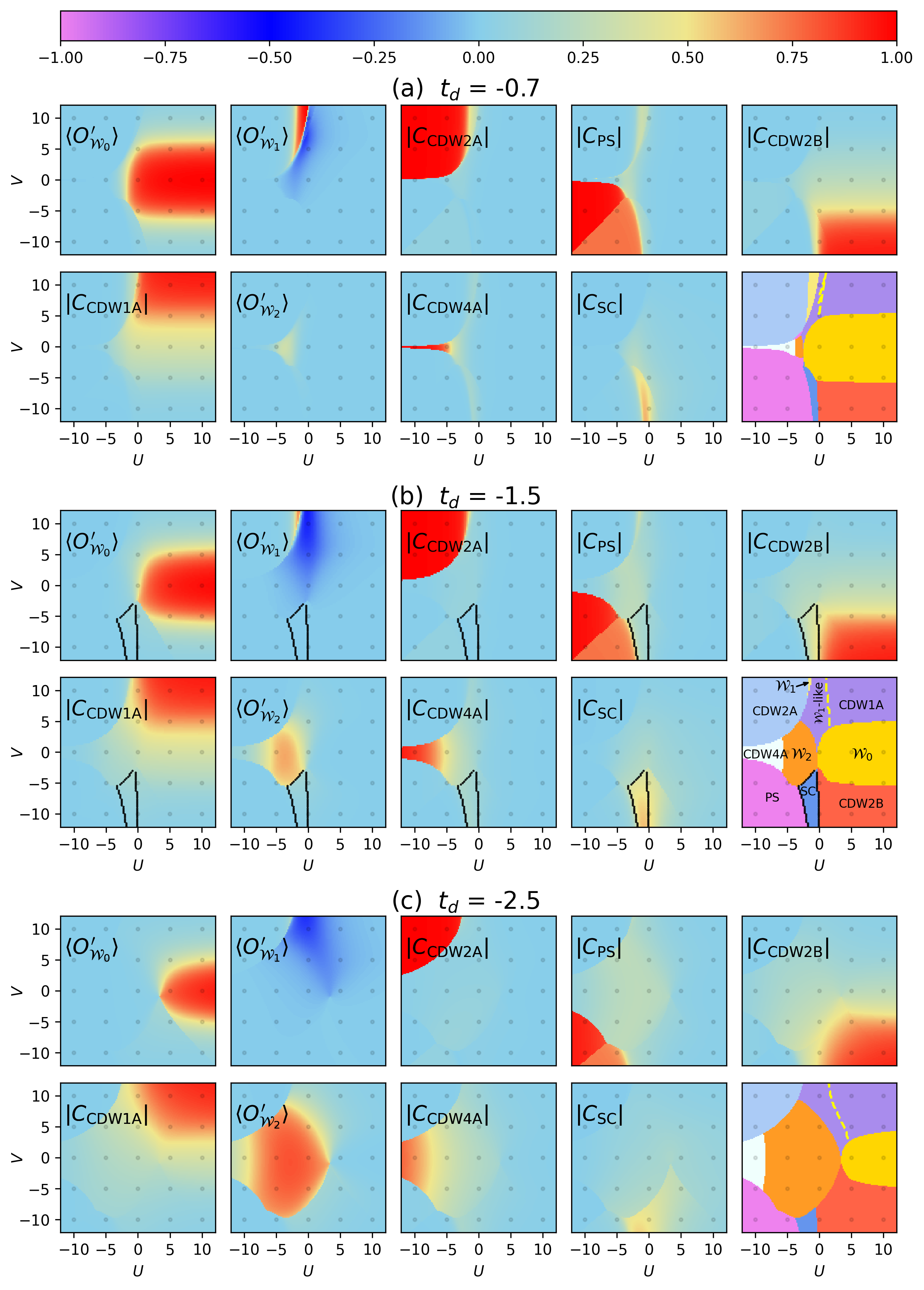}
\caption{
%
%
%
%
The OPs for the possible phases in the interacting ESSH model with various $t_d$: (a) $t_d=-0.7$; (b) $t_d=-1.5$; and (c) $t_d=-2.5$. Here, $N=8$, $(\eta, t_c)=(0.6,0)$. $O_{\mathcal{W}_{0/1/2}}'$ are defined in Eq. (\ref{eqn:wm_combination}), with $n=3$. $C_\text{m}(r)$ is defined in Eq. (\ref{eqn:correlation_function}), and ``m'' can be CDW1A/2A/2B/4A or SC or PS. For the CDW family, $r$ is set as halve of the wavelength of the corresponding phase, namely $r=1$ for CDW2A/2B and $r=N/2$ for PS. For CDW1A, $r=0.5$, which corresponds to a separation of one site. For the SC-like phase, we set $r=1$ and  $O_{\text{SC}}$ is given in Eq. (\ref{eqn:sc_pairing_op}). The bottom-right panel in each subplot shows the estimated phase diagram. The yellow dashed line display the boundary between the $\mathcal{W}_1$-like and the CDW-1A phases approximated from the peak of the derivative of $\langle O_{\mathcal{W}_1}'\rangle $.} 
  \label{fig:bigbig}
\end{figure*}


\subsubsection{Key features of the OPs in \\capturing the phase diagram}
\label{subsec:key_features}
In this subsection, we present the ground-state phase diagrams for small systems. Under PBC, the ground states of the CDW-1A/2A/2B/4A and the PS phases are symmetry-unbroken in finite systems. Therefore, instead of using the OP expectation values themselves, we consider their correlation functions: 
\begin{align}
    C_{\text{m}}(r)= \langle O_\text{m}^\dag(i+r)O_\text{m}(i)\rangle -  \langle O_\text{m}^\dag(i+r)\rangle \langle O_\text{m}(i)\rangle, 
    \label{eqn:correlation_function}
\end{align} 
where m denotes CDW1A/2A/2B/4A or PS, and the operators $O_\text{m}(j)$ are defined in Eqs. (\ref{op:cdw1a})/ (\ref{op:cdw2a})/ (\ref{op:cdw2b})/ (\ref{op:cdw4}) or (\ref{op:ps}), respectively. For brevity, we sometimes refer to the correlation function associated with the designated OP simply as ``the OP" in what follows, especially in the context of numerical simulations.

Figure \ref{fig:bigbig} shows the OPs of the possible phases on the $U$-$V$ plane with $(\eta,t_c) = (0.6,0)$ for $t_d=-0.7$, $t_d=-1.5$, and $t_d=-2.5$. The results are obtained from exact diagonalization (ED). 
The case of $t_c\neq0$ and $\eta=\pm0.6$ are discussed in Appendix \ref{app:big_appendix_interacting} for completeness. For the moment, let us first ignore the subplots labeled as $|C_{SC}|$, which will be discussed in the subsection that follows. The OPs attain finite values in their own phases, in agreement with the schematic phase diagram shown in Fig. \ref{fig:schematic_phasediagram}, and vanish in other phases. Note that the range of $U$ and $V$ are extended here, as compared to that in Fig. \ref{fig:builidingInteractingOp}, to capture the CDW-1A and CDW-2B phases above and below the $\mathcal{W}_0$ phase, respectively. The bottom-right panel in each subplot displays the estimated phase diagram by comparing the magnitudes of the OPs. The phase is assigned according to the corresponding OP that has the largest magnitude.

If one considers the plot of $|C_{\rm{PS}}|$, it appears to have two phases separated by a sharp change of $|C_{\rm{PS}}|$ in the PS region. However, this is just a finite-size effect. The boundary is only present in small systems with even number of unit cells, and is absent in systems with odd number of unit cells. Detailed discussion will be provided in Sec. \ref{subsec:PS_sizeeffect}.

As $t_d$ decreases from -0.7 to -2.5, we also observed that the $\mathcal{W}_2$ and the CDW-4A phases expand. The former is intuitive to understand as we have emphasized throughout the paper that the $\mathcal{W}_2$ phase is favored by $t_d$ hoppings. On the other hand, the latter is consistent with the claim in earlier section that the emergence of the CDW-4A phase is a consequence of the interplay between beyond nearest-neighbor hoppings and two-body interactions.

Furthermore, we observe a non-negligible negative value in $\langle O_{\mathcal{W}_1}'\rangle$. One may immediately associate this with the $\mathcal{W}_1^p$ phase, discussed in \cite{hui2026}, since the $\mathcal{W}_1^p$ phase is characterized by a negative value of the OP in the non-interacting case as defined in Eq. (\ref{eqn:w1_non-interacting}) (the OP ``normally'' gives positive value). 
The appearance of this phase is a consequence of introducing the further neighbor electron-like $t_d$ hopping ($t_d<0$), which is absent in Fig. \ref{fig:builidingInteractingOp}(b) even if one extends the ranges of $U$ and $V$. However, as we will see later in Sec. \ref{sec:DMRG_result}, this is not truly a $\mathcal{W}_1^p$ phase. To distinguish it, we refer to it as a $\mathcal{W}_1$-like phase, whose OP is yet to be determined. 


Focusing on Fig. \ref{fig:bigbig}(b), and again disregarding the $|C_{\text{SC}}|$ plot, a cautious reader may notice there is a region, enclosed by the black dashed lines, that is not captured by any of the other OPs. This region turns out to be a novel phase that we have not yet discussed. It arises from the introduction of further neighbor hopping $t_d$ and interactions. In what follows, we argue that this is a superconducting-like (SC-like) phase. 

\subsubsection{The superconducting-like phase}
To find the OP of this phase, we consider $(\eta, t_c, t_d,U,V)=(0.6, 0,-1.5,-1,-12)$ as an example. The dominant Fock states of the ground state which has $|\xi_i|^2>0.005$ are:  
\begin{eqnarray}
\label{Fock:omg1}
&|\widehat{1 \ 0}\widehat{0\ 1}\widehat{1\ 0}\widehat{0\ 1}\widehat{1\ 0}\widehat{0\ 1}\widehat{1\ 0}\widehat{0\ 1}\rangle \ \ \  |\xi_i|=0.17720,\\
\label{Fock:omg2}
&|\widehat{1 \ \underline{0}} \widehat{\underline{0}\ \underline{0}} \widehat{\underline{0}\ \underline{1}} \widehat{\underline{1}\ \underline{1}} \widehat{\underline{1}\ 0}\widehat{0\ 1}\widehat{1\ 0}\widehat{0\ 1}\rangle \ \ \ |\xi_i|=0.13299,\\
\label{Fock:omg3}
&|\widehat{1 \ \underline{0}} \widehat{\underline{0}\ \underline{0} }\widehat{\underline{0}\ 1}\widehat{1\ 0}\widehat{0\ \underline{1}} \widehat{\underline{1}\ \underline{1}} \widehat{\underline{1}\ 0}\widehat{0\ 1}\rangle \ \ \ |\xi_i|=0.12500,\\
\label{Fock:omg4}
&|\widehat{\underline{1} \ \underline{0}} \widehat{\underline{0}\ \underline{0} }\widehat{\underline{0}\ \underline{1} }\widehat{\underline{1}\ \underline{1} }\widehat{\underline{1}\ \underline{0} }\widehat{\underline{0}\ \underline{0} }\widehat{\underline{0}\ \underline{1} }\widehat{\underline{1}\ \underline{1}}\rangle \ \ \ |\xi_i|=0.11343,
\end{eqnarray} 
where Eqs. (\ref{Fock:omg1}), (\ref{Fock:omg2}),(\ref{Fock:omg3}), (\ref{Fock:omg4}), represent families of 2, 16, 8, 4 different Fock states, respectively, which are equivalent up to a global translation of one unit cell. Note that Eq. (\ref{Fock:omg1}) corresponds to the configuration that also dominates in the CDW-2B phase, while Eqs (\ref{Fock:omg2})-(\ref{Fock:omg4}) deviate from it by a pair of CDW-4B configurations, as underlined. 

To relate the Fock states in Eq. (\ref{Fock:omg1}) to Eq.(\ref{Fock:omg2}) by an operator, a suitable choice is $c^{\dag}_{3,B}c^{}_{2,B} c^{\dag}_{4,A}c^{}_{3,A}$, or more generally, 
\begin{equation}
    \label{eqn:omg_initial}
    \begin{split}
          C_{\text{SC}} &= 2\langle c^{\dag}_{j+1,B}c^{}_{j,B} c^{\dag}_{j+2,A}c^{}_{j+1,A}\rangle ,
    \end{split}
\end{equation}
where the factor of 2 is just a normalization constant. The consistency of this notation with Eq. (\ref{eqn:correlation_function}) is shown later. The bottom-right panel in Figs. \ref{fig:bigbig}(a)-(c) showcase the results of $|C_{\text{SC}}|$ in Eq. (\ref{eqn:omg_initial}), where the missing phase enclosed by the black dashed line is perfectly captured. 

\begin{figure}
  \includegraphics[width=8cm]{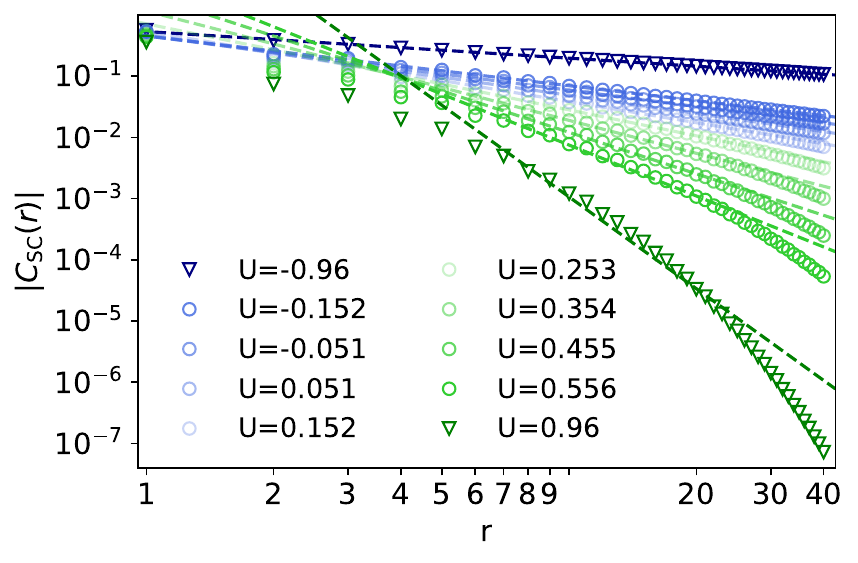}
  \caption{The log-log plot of the pairing correlation function in Eq. (\ref{eqn:sc_correlation2}) versus the separation $r$ measured from the middle of the chain. Blue and green markers denote the SC-like and the CDW-2B phases, respectively, with $U$ values indicated in the legend and $(\eta, t_c,t_d,V)=(0.6,0,-1.5,-12)$. Triangular and circular markers denote points deep inside the phase and around the phase boundary, respectively. Data is obtained from DMRG simulation with $N=200$, open boundary condition, bond dimension 500, and at least 10 sweeps. Dashed straight lines are linear fits to the data over $r\in[6,25]$.}
  \label{fig:correlation_sc}
\end{figure}

Physically, Eq. (\ref{eqn:omg_initial}) moves two spinless fermions from the $(j+1,A)$ and the $(j,B)$ sites to the $(j+2,A)$ and the $(j+1,B)$ sites. The hopping always involves a pair of spinless fermions. If we introduce an additional hopping operator to Eq. (\ref{eqn:omg_initial}), for example, $c^{\dag}_{j+3,A}c^{}_{j+2,A}c^{\dag}_{j+1,B}c^{}_{j,B} c^{\dag}_{j+1,A}c^{}_{j,A}$, its expectation value always vanishes in this phase. This in fact holds for operators with odd number of hopping terms. This observation motivates us to name this phase as an SC-like phase, analogous to the Cooper pairs in superconductivity.

While conventional superconductivity involves Cooper pairs of opposite spins, the sublattice degree of freedom of the current system can be loosely viewed as playing a role similar to spin. The key insight is that 1D spinless fermions can exhibit an SC-like behavior in the presence of extended hopping ($t_d\neq 0$) and imbalance of attractive potential between neighboring sites ($U\neq V$). 

\begin{figure*}
    \centering
    \includegraphics[width=16cm]{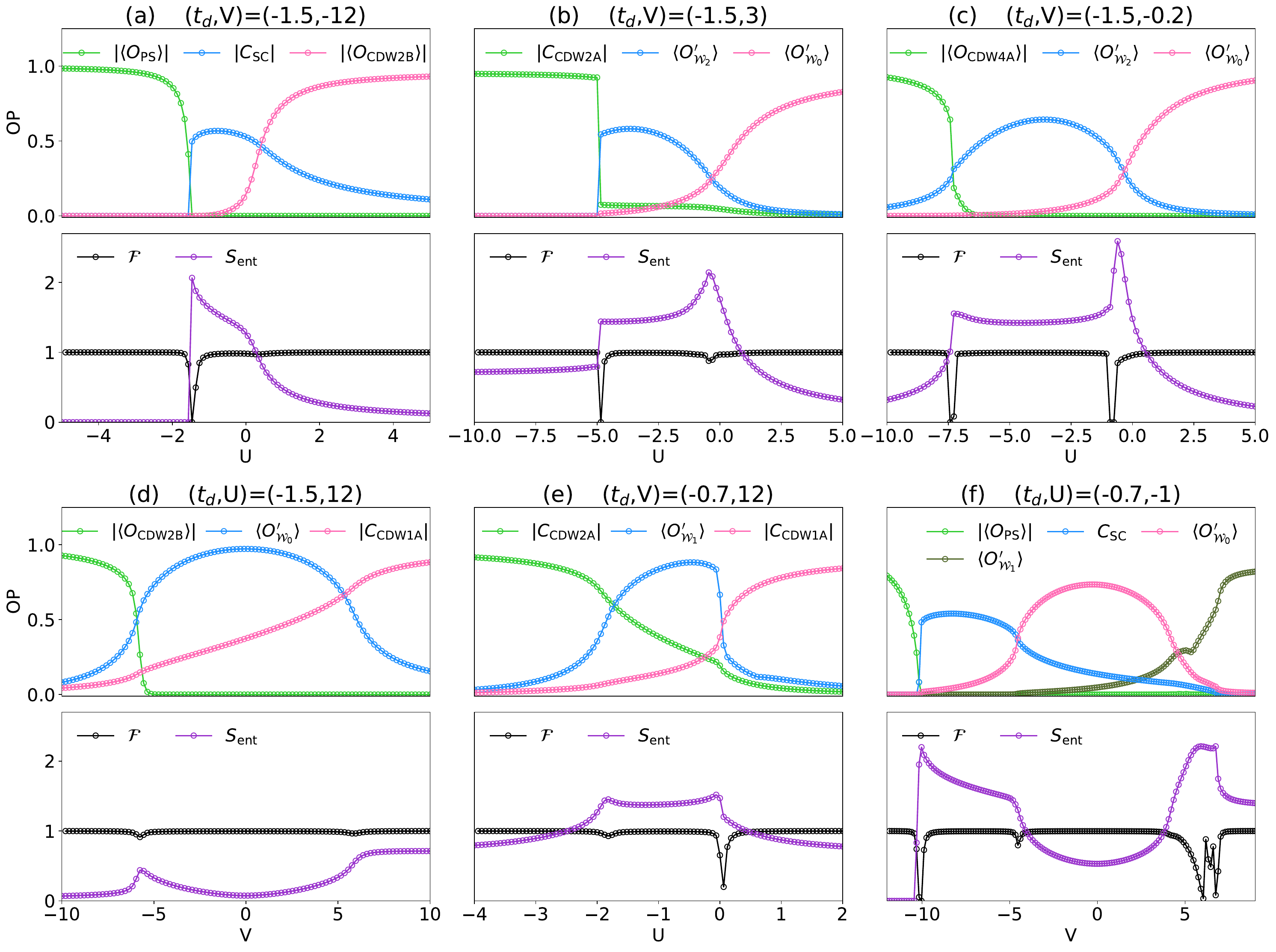}
    \caption{The OPs (top panel of each subplots), the half-block entanglement entropy $S_{ent}$ and the fidelity $\mathcal{F}$ (bottom panel of each subplots) along
    six chosen paths in Fig. \ref{fig:bigbig} (a) and (b) for $N=90$ and $(\eta,t_c)=(0.6,0)$. The six paths have included all the phases discussed in Sec. \ref{sec:OP_interacting}. 
    The OP is the largest among the others in its respective phase, and the crossings agree well with the phase transition detected by the entanglement entropy and the fidelity, except for the $\mathcal{W}_1$-like phase in (f).}
    \label{fig:dmrg}
\end{figure*}

To further the analogy, we reorder the creation and annihilation operators in Eq. (\ref{eqn:omg_initial}) and generalize it to 
\begin{align}
    \label{eqn:omg_correlation1}
    C_{\text{SC}}(r) = 2\langle c^{\dag}_{j+1+r,A}c^{\dag}_{j+r,B}c^{}_{j,B} c^{}_{j+1,A}\rangle.
\end{align}
The above is reduced to Eq. (\ref{eqn:omg_initial}) for $r=1$. 
Defining 
\begin{align}
    O_\textrm{SC} (j)\equiv \Delta^\dag_j = c^{\dag}_{j+1,A}c^{\dag}_{j,B},
    \label{eqn:sc_pairing_op}
\end{align}
Eq. (\ref{eqn:omg_correlation1}) can be rewritten as
\begin{align}
    C_{\text{SC}}(r) &= 2\langle \Delta_{j+r}^\dag \Delta^{}_{j}\rangle \nonumber\\
    &=2\Bigl(\langle\Delta_{j+r}^\dag \Delta^{}_{j}\rangle - \langle\Delta_{j+r}^\dag \rangle\langle\Delta^{}_{j}\rangle \Bigl).
    \label{eqn:sc_correlation2}
\end{align}
This can be viewed as the correlation function in Eq. (\ref{eqn:correlation_function}) of the pairing order parameter $\Delta^\dag$, analogous to superconductivity. The last equality can be justified by noticing that we restrict the ED calculation in half-filling, where $\langle c^\dag_ic^\dag_j\rangle=0\  \forall i,j$. Thus, $\langle\Delta^\dag_{j+r}\rangle=0$ and $\langle\Delta^{}_j\rangle=0$.

By the Mermin–Wagner theorem, true long-range superconducting order cannot exist in strictly 1D systems with short-range interactions. However, quasi-long-range order with algebraic decay correlations is allowed. Figure \ref{fig:correlation_sc} shows the log-log plot of $| C_{\text{SC}}(r)|$ in Eq. 
(\ref{eqn:sc_correlation2}) as a function of the separation distance $r$. Deep inside the SC-like phase, the correlation function follows a power law decay in $r$, evidenced by the straight line in the log-log plot (the dark blue triangles). This suggests the existence of quasi-long-range order. As the system transition from the SC-like to the CDW-2B phase (blue to green circles), the correlation function no longer follows a power-law decay with distance. This deviation is more pronounced when the system is deep in the CDW-2B phase (dark green triangles).  

We remark that there is another type of SC-like phase emerge when considering $t_c\neq0$ and $t_d=0$, which is discussed in Appendix \ref{app:big_appendix_interacting}.

\begin{table}[]
\begin{tabular}{|c|cc|}
\hline
                   & \multicolumn{1}{c|}{$(U,V)=(-10,-5)$}                       & $(U,V)=(-5,-10)$                       \\ \hline
Even   & \multicolumn{1}{c|}{$|\widehat{11}\ \widehat{00}\ \widehat{00}\ \widehat{00}\ \widehat{11}\ \widehat{11}\rangle$}                   & $|\widehat{10}\ \widehat{00}\ \widehat{00}\ \widehat{01}\ \widehat{11}\ \widehat{11}\rangle$                   \\ \hline
Odd & \multicolumn{2}{c|}{\begin{tabular}[c]{@{}c@{}}$|\widehat{11}\ \widehat{00}\ \widehat{00}\ \widehat{01}\ \widehat{11}\rangle$\\ $|\widehat{10}\ \widehat{00}\ \widehat{00}\ \widehat{11}\ \widehat{11}\rangle$\end{tabular}} \\ \hline
\end{tabular}
\caption{The dominant Fock states for the ground states in the PS phase for $U<V<0$ and $V<U<0$, with
$(\eta,t_c,t_d)=(0.6,0,0)$. The $U$ and $V$ values are specified in the first row. Each Fock state represents a family of Fock states that are equivalent under a global translation by \emph{one unit cell} and a Hermitian conjugate. For illustration, system size of $N=6$ (even) and $N=5$ (odd) are considered.}
\label{tab:ps_finitesizeeffect}
\end{table}

\begin{figure*}
    \centering
    \includegraphics[width=16cm]{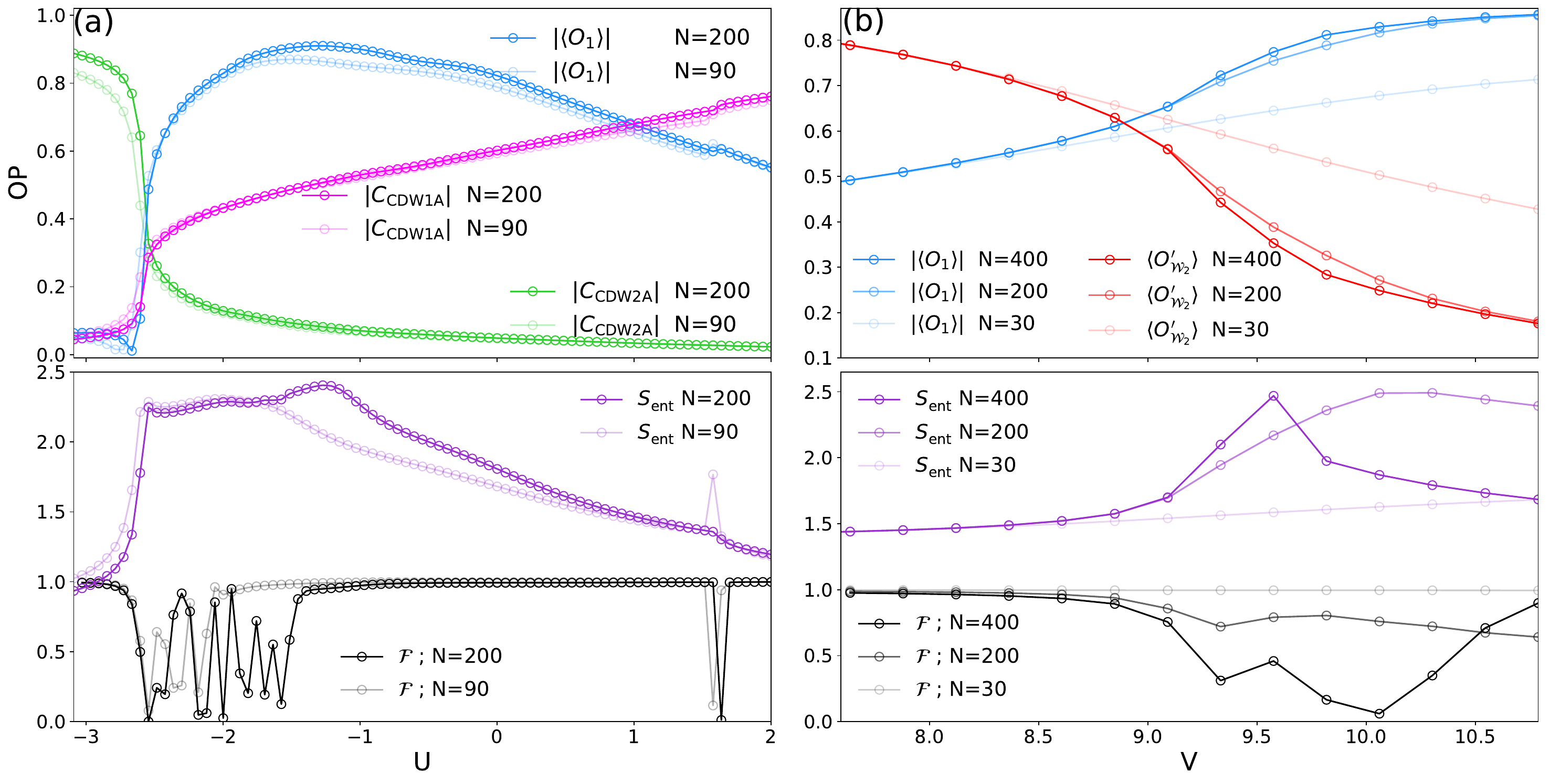}       
    \caption{The OPs (top panel), half-block entanglement entropy $S_{ent}$ and fidelity $\mathcal{F}$ (bottom panel) along two selected paths in Fig. \ref{fig:bigbig}(c) that enclose the $\mathcal{W}_1$-like phase with various system sizes. The parameters are fixed as (a) $(\eta, t_c, t_d,V)=(0.6, 0, -2.5,12)$ and (b) $(\eta, t_c,t_d,U)=(0.6,0,-2.5,-2)$. 
    The smaller the system size is, the more transparent the markers are.}
    \label{fig:dmrg_finiteSizeEffect}
\end{figure*}

\subsubsection{The finite-size effect in the PS phase}
\label{subsec:PS_sizeeffect}
As noted in Sec. \ref{subsec:key_features}, the PS phase exhibits a clear change in $|C_{\text{PS}}|$ within its own phase in Fig. \ref{fig:bigbig}, which we attribute to finite-size effect. The ground state's dominant Fock states in each region are summarized in Table \ref{tab:ps_finitesizeeffect}. For systems containing even number of unit cells, the dominant Fock states at $(U,V)=(-10,-5)$ and $(-5,-10)$ are slightly different, unlike in the case of odd number of unit cells. This slight difference explains the apparent boundary inside the PS phase.

A natural question to ask is why there exists such a difference in systems with even $N$ but not odd $N$. The essence lies in the minimization of energy. In systems with even $N=6$, the dominant Fock state in the left column in Table \ref{tab:ps_finitesizeeffect} contributes an energy of approximately $3U+2V$. On the other hand, the dominant Fock state in the right column contributes an energy of approximately $2U+3V$. Therefore, $\ket{11\ 00\ 00\ 00\ 11\ 11}$ is preferred for $U<V<0$ while $\ket{10\ 00\ 00\ 01\ 11\ 11}$ is preferred for $V<U<0$ in terms of energy minimization. On the other hand, both $\ket{11\ 00\ 00\ 01\ 11}$ and $\ket{10\ 00\ 00\ 11\ 11}$ contribute the same energy of $2U+2V$ in the odd $N=5$ case, and the finite-size boundary in the PS phase is absent. 
In large systems, the energy difference becomes insignificant and finite-size effect is negligible, which we will further verify with results from DMRG in the following subsection.

\subsection{OPs along certain paths in large systems}
\label{sec:DMRG_result}

Figure \ref{fig:dmrg} shows the OPs along six selected paths in the $U$-$V$ phase diagram for $t_d=-0.7$ and $t_d=-1.5$, which include all the phases discussed above. The results are obtained from DMRG simulation of a system of $N=90$ unit cells with open boundary condition (OBC). Depending on the phase, 300-1000 Schmidt states are kept and 10-25 sweeps are performed in the simulation. The OPs are evaluated at the middle of the chain to reduce edge effects. For the CDW-1A/2A and the SC-like phases, the respective OP is taken as the correlation function in Eq. (\ref{eqn:correlation_function}), similar to the ED case. For the CDW-2B/4A and PS phases, we consider the OP itself (i.e. $\langle O_\text{m}(j)\rangle$) instead of the correlation function. Since under OBC and an even number of unit cells, the ground states in these phases are symmetry-broken (see Appendix \ref{app:obc_ED}).

As shown in the figure, the OP is the largest among the others in its own respective phase, and the crossings between the respective OPs (top panels in each subplot) with the others agree qualitatively well with the transitions predicted by the entanglement entropy and fidelity (bottom panel in each subplot), except for the case involving $O_{\mathcal{W}_1}'$ in Fig. \ref{fig:dmrg}(f).

For the path in Fig. \ref{fig:dmrg}(f), recall that for the $N=8$ result shown in Fig. \ref{fig:bigbig}(a), $\langle O_{\mathcal{W}_1}'\rangle$ is negative in the ${\mathcal{W}_1}$-like phase. Such a negative value also appears in other small-size systems and is in fact a finite-size effect. In large systems, $\langle O_{\mathcal{W}_1}'\rangle$ is positive even in the $\mathcal{W}_1$-like phase (Fig. \ref{fig:dmrg}(f)). This suggests that the $\mathcal{W}_1$-like phase is not exactly a $\mathcal{W}_1^p$ phase. 
Furthermore, the fidelity exhibits strong fluctuation in the $\mathcal{W}_1$-like phase. Such a strong fluctuation in the fidelity usually indicates the ground state is degenerated \cite{ali2021}, unlike the $\mathcal{W}_1$ phase, in which the ground state is non-degenerated. Any linear combination within the degenerated subspace is also a valid ground state. In each DMRG calculation, the program could end up with significantly different linear combinations and thus results in fluctuating values of the fidelity.


Moreover, the entanglement entropy clearly indicates a transition between the $\mathcal{W}_1$ and the $\mathcal{W}_1$-like phases around $V=6.8$ in Fig. \ref{fig:dmrg}(f), suggesting the two phases are distinct. Thus, we might apply the scheme in Ref. \cite{hui2026} to determine a trial OP of the $\mathcal{W}_1$-like phase, which reads $O_{1}(j) = 2(
c_{j+1,A}^\dag c_{j+1,B}^{} c_{j+2,A}^\dag c_{j,B}^{}
+\text{h.c.})$.
However, as we demonstrate below, this is not a perfect OP due to strong finite-size effect of the phase concerned, which marks the limitation of the scheme. 

The OPs derived from the scheme rely on ground states for system size accessible by ED. This implicitly assumed that small systems exhibit behavior that is qualitatively similar to that of large systems. If small systems differ markedly from large ones, it is natural to expect the scheme to fail. Indeed, our numerical simulations on large systems indicate significant finite-size effect in the $\mathcal{W}_1$-like phase. As the system size increases, the degenerated region expands with no clear indication of convergence. This can be seen from Fig. \ref{fig:dmrg_finiteSizeEffect}(a), which shows the OPs (top panel), the half-block entanglement entropy and the fidelity (bottom panel) for $(\eta,t_c,t_d,V) = (0.6,0,-2.5,12)$ in systems of $N=90$ and $N=200$. The behavior of the entanglement entropy and the fidelity are inconsistent across various system sizes. While fluctuations in the fidelity can be attributed to the ground-state degeneracy, the degenerate region expands as the system size increases, underscoring the importance of finite-size effects. A similar conclusion follows from Fig. \ref{fig:dmrg_finiteSizeEffect}(b), which shows stark differences among system sizes $N=30, 200, 400$.


\section{Conclusion}
\label{sec:conclusion}

We investigate the ESSH model with intra-cell and inter-cell interactions. We first explore the emergence of various phases under different limits of the hopping and interaction parameters. In one of the limits, we elucidate that interactions do not necessarily destroy the topological structure of the ground state. Additionally, five different types of CDW phases are identified, two of which arise from the interplay between imbalanced interactions and extended hoppings. Moreover, a superconducting-like phase is discovered in this model, which has not been reported previously, to the best of our knowledge. The SC-like phase exhibit quasi-long-range order, as expected from the Mermin-Wagner theorem. 

Applying the scheme proposed in Ref. \cite{hui2026}, we identify the corresponding OPs capturing all the topological and the trivial phases in the interacting model. Unlike the non-interacting case, where the number of hoppings in the OPs can be chosen as half the chain, interactions restrict the allowable number of hoppings, according to our analysis. Therefore, the OPs have to be refined to account for the effect of interactions, which facilitates non-unidirectional hoppings in the dominant Fock states. The OPs we construct yield consistent results in large-system analysis, in agreement with diagnostics including the entanglement entropy and fidelity, except for the $\mathcal{W}_1$-like phase that suffers from strong finite-size effect. This exposes a limitation of our approach that is unobvious in the non-interacting case in Ref. \cite{hui2026}: it may fail for phases whose features differ substantially between small and large systems.

Regarding future works, the properties of the novel phases we find in this study, for example the SC-like phases and the $\mathcal{W}_1$-like (as well as the $\mathcal{W}_0$-like) phase, merit further investigations. One potential diagnostics could be the topological markers \cite{chen2023,melo2023}.

\emph{Note added. } After the initial submission of this work, we became aware of an independent concurrent work reported in Ref. \cite{niu2026}. Although the authors in Ref. \cite{niu2026} also considered an interacting ESSH model, the phase diagram considered here is different from that in their work. Specifically, our analysis incorporates additional model parameters, which reveals new physical features of the model. While Ref. \cite{niu2026} studies the topological phases using the well-known string order parameter, here we derive the OPs which provide physical insights into the ground-state's dominant configurations and further differentiate between phases with different winding numbers. Additionally, we elucidate that interactions do not necessarily destroy the topological structure of the ground state. Taken together, Ref. \cite{niu2026} and our work offer complementary perspectives that contribute to a broader understanding of the interacting ESSH model.

\begin{acknowledgments}
We acknowledge financial support from Research Grants Council  of  Hong  Kong  (Grant  No. CityU 11318722),  
CeFEMA, Centre of Physics and
Engineering of Advanced Materials (under contracts
UID/04540/2025, UID/PRR/04540/2025 and UID/PRR2/04540/2025 with the
Portuguese Agência para a Investigação e Inovação AI2), LaPMET, Laboratory of Physics for
Materials and Emerging Technologies Portuguese (under
contract LA/P/0095/2020 with the Portuguese Agência para a Investigação e
Inovação AI2),
and City University of Hong Kong (Grants No. 9610438, 7020156). PDS thanks the hospitality during his visit to City University of Hong Kong. 
\end{acknowledgments}

\appendix
\titlelabel{\ifcsname c@section\endcsname Appendix~\thesection:\ \fi}

\section{Additional phase diagrams in the\\ interacting ESSH model}
\label{app:big_appendix_interacting}
\begin{figure*}
\includegraphics[width=16.8cm]{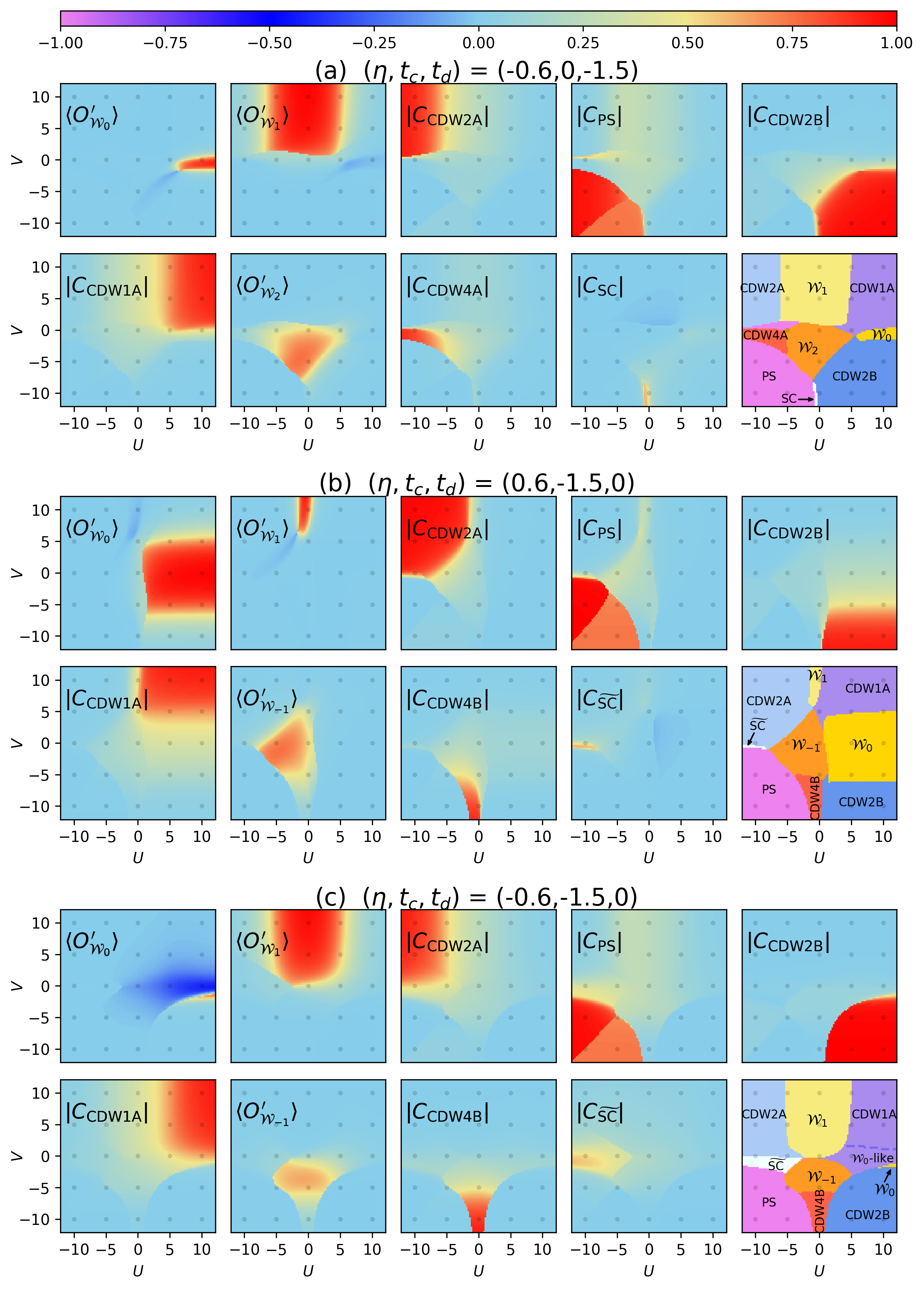}
\caption{
The OPs for the possible phases in the interacting SSH model with $(\eta,t_c,t_d)$: (a) $(-0.6,0,-1.5)$; (b) $(0.6,-1.5,0)$; and (c) $(-0.6,-1.5,0)$. $O_{\mathcal{W}_{-1}}'$ and $C_{\widetilde{SC}}$ is defined in Eq. (\ref{eqn:wm_combination}) and Eq. (\ref{eqn:sc_tc-type_correlation}), respectively. The definition of other quantities are the same as Fig. \ref{fig:bigbig}. The bottom-right panel in each subplot shows the estimated phase diagram. The purple dashed line display the boundary between the $\mathcal{W}_0$-like and the CDW-1A phases approximated from the peak of the derivative of $\langle O_{\mathcal{W}_0}'\rangle$.
}
\label{fig:bigappendix}
\end{figure*}
Figure \ref{fig:bigappendix} shows the ground-state phase diagrams for three additional cases with different choices of $\eta,t_c,t_d$. All the three cases in Fig. \ref{fig:bigappendix} host the CDW-2A, CDW-1A, CDW-2B, and PS phases residing at the expected corners in the phase diagram, consistent with the schematic phase diagram in Fig. \ref{fig:schematic_phasediagram}.

In Fig. \ref{fig:bigappendix}(a), we consider the case $(\eta,t_d)=(-0.6,-1.5)$ with $t_c=0$. The phase diagram can be compared with that in Fig. 7(b) of Ref. \cite{yu2016} where $t_c=t_d=0$, or with that in Fig. \ref{fig:bigbig}(b) in the main text where $\eta=0.6$. The $\mathcal{W}_0$ ($\mathcal{W}_{1}$) phase emerges at sufficiently large $U$ (large $V$) with $V\approx0$ ($U\approx0$), as expected from the discussion in Sec. \ref{sec:interacting_limit_deep_w0_w1}. There is a small region above the CDW-4A phase being identified as a PS phase. However, this is in fact a finite-size effect. Analysis of large-system simulation data (not presented here) show that it is a shallow region in the $\mathcal{W}_2$ phase.

In Fig. \ref{fig:bigappendix}(b), we consider the case  $(\eta,t_c)=(0.6,-1.5)$ with $t_d=0$. The presence of the $t_c$ hopping allows the  appearance of the $\mathcal{W}_{-1}$ phase along with the CDW-4B phase, as argued in Sec. \ref{sec:limiting_case_CDW4A4B}. Similar to Fig. \ref{fig:bigappendix}(a), a small region to the right of the CDW-4B phase in the estimated phase digram is misidentified as a CDW-2B phase due to finite-size effects. It is expected to be a shallow $\mathcal{W}_{-1}$ phase instead.

In the main text, we demonstrate that an SC-like phase appears when the extended-range $t_d$ hopping is introduced. Similarly, another SC-like phase arises when we instead include the other type of extended-range hopping, $t_c$. We denote this phase by $\widetilde{\text{SC}}$ in the phase diagram [Fig. \ref{fig:bigappendix}(b)] to distinguish it from the SC-like phase discussed in the main text. The OP of this $\widetilde{\text{SC}}$-like phase is:
\begin{align}
    O_{\widetilde{\text{SC}}}(j)\equiv\widetilde{\Delta}_j^\dag = c_{j,A}^\dag c_{j,B}^\dag\ ,
\end{align}
which is shifted by one site relative to Eq. (\ref{eqn:sc_pairing_op}). Consequently, we can write down: 
\begin{align}
    C_{\widetilde{\text{SC}}}(r) &= 2\langle c^{\dag}_{j+r,A}c^{}_{j,A} c^{\dag}_{j+r,B}c^{}_{j,B} \rangle=2\langle\widetilde{\Delta}_{j+r}^\dag\widetilde{\Delta}_j^{}\rangle \ .
    \label{eqn:sc_tc-type_correlation}
\end{align}

Interestingly, the phase diagrams in Fig. \ref{fig:bigappendix}(a) and (b) appear to be related by a reflection about the $U=V$ line, with the phases mapped accordingly. We can understand this as follows: The reflection corresponds to $(U,V)\mapsto(V,U)$. Additionally, notice that the parameters between (a) and (b) transform as $(t_a,t_b,t_c,t_d)\mapsto(t_b,t_a,t_d,t_c)$. Therefore, one may identify the site maps as $(j,A)\mapsto(j,B)$ and $(j,B)\mapsto(j+1,A)$. Under this mapping, the ``$A$" and ``$B$'' sublattice labels are exchanged in all terms (both the kinetic and the interaction terms) of the Hamiltonian in Eq. (\ref{eqn:ISSH}).

In Fig. \ref{fig:bigappendix}(c), we consider $(\eta,t_c)=(-0.6,-1.5)$ with $t_d=0$. Similar to Fig. \ref{fig:bigappendix}(b), the $\widetilde{\text{SC}}$-like and the CDW-4B phases appear. We conjecture that there is also a counterpart of the $\mathcal{W}_1$-like phase, which we refer to as the $\mathcal{W}_0$-like phase, as hinted by the negative value of $\langle O'_{\mathcal{W}_0}\rangle$. However, as with the $\mathcal{W}_1$-like phase, its OP remains to be identified. In parallel with the the mapping $(U,V)\mapsto(V,U)$ between Fig. \ref{fig:bigappendix}(a) and (b), Fig. \ref{fig:bigappendix}(c) also exhibits similar mapping with Fig. \ref{fig:bigbig}(b).

\section{Numerical justification of symmetry breaking in certain phases under OBC}
\label{app:obc_ED}

Consider $(\eta,t_c,t_d,U,V)=(0.6,0,0,12,-10)$, which is a CDW-2B phase. Calculating the ground state by ED under OBC, we find that the dominant Fock states reads:
\begin{align}
    |\widehat{0 \ 1}\widehat{1\ 0}\widehat{0 \ 1}\widehat{1\ 0}\widehat{0 \ 1}\widehat{1\ 0}\widehat{0 \ 1}\widehat{1\ 0}\rangle \ \ \ |\xi_i|=0.8923,
    \label{fock:obc_cdw2b}
\end{align}
which does \emph{not} include the counterpart state $|\widehat{1 \ 0}\widehat{0\ 1}\widehat{1 \ 0}\widehat{0\ 1}\widehat{1 \ 0}\widehat{0\ 1}\widehat{1 \ 0}\widehat{0\ 1}\rangle$, unlike the PBC case in Sec. \ref{sec:limitingcase_CDW2B}. This counterpart Fock state contributes around $3V$ of energy, whereas the Fock state in Eq. (\ref{fock:obc_cdw2b}) contributes an energy of around $4V$. Therefore, the Fock state in Eq. (\ref{fock:obc_cdw2b}), which has a lower energy for negative $V$, is preferred by the ground state. In contrast, both Fock states contribute an energy of approximately $4V$ under PBC, resulting in equal weighting in the ground state. Thus, the symmetry broken ground state in DMRG calculation in the CDW-2B phase is due to the boundary condition, justifying the use of $\langle O_{\text{CDW2B}}(j)\rangle$.

Consider $(\eta,t_c,t_d,U,V)=(0.6,0,0,-12,-10)$, which is a PS phase. The dominant Fock state in the case of OBC is: 
\begin{align}
    |\widehat{00}\ \widehat{00}\ \widehat{11}\ \widehat{11}\ \widehat{11}\ \widehat{11}\ \widehat{00}\ \widehat{00} \rangle \ \ \ |\xi_i|=0.9988,
    \label{fock:obc_ps}
\end{align}
and this also does not include the counterpart state equivalent up to a global translation of one unit cells like the one in Sec. \ref{sec:limitingcase_ps} in the case of PBC. The reason for resulting in such a symmetry broken ground state is similar to the case of the CDW-2B phase above.

For $(\eta,t_c,t_d,U,V)=(0.6,0,-1.5,-12,-0.2)$, which is a CDW-4A phase. With a similar reason as the above cases, the only dominant Fock state in the ground is 
\begin{align}
    |\widehat{00}\ \widehat{11}\ \widehat{11}\ \widehat{00}\ \widehat{00}\ \widehat{11}\ \widehat{11}\ \widehat{00} \rangle \ \ \ |\xi_i|=0.9406,
    \label{fock:obc_cdw4a}
\end{align}
instead of having a counterpart state that is equivalent up to a global translation of one unit cells like the one in Sec. \ref{sec:limiting_case_CDW4A4B}. However, note that for $V=0$, the ground state of the CDW-4A phase is symmetry unbroken.

\bibliography{bibfile}

\end{document}